\documentclass[fleqn,usenatbib]{mnras}

\usepackage{newtxtext,newtxmath}
\usepackage[T1]{fontenc}
\usepackage{ae,aecompl}
\usepackage{graphicx}
\usepackage{amsmath}
\usepackage{amssymb}
\usepackage{natbib}
\usepackage{gensymb}
\usepackage{upgreek}
\usepackage{comment}

\title[High precision pulsar timing and spin frequency second derivatives]{High precision pulsar timing and spin frequency second derivatives}

\author[X.~J.~Liu, C.~G.~Bassa \& B.~W.~Stappers]{   
	X.~J.~Liu$^{1}$\thanks{E-mail:xiao-jin.liu@postgrad.manchester.ac.uk},
	C.~G.~Bassa$^{2}$,
	B.~W.~Stappers$^{1}$
    \\
	$^{1}$Jodrell Bank Centre for Astrophysics, School of Physics and Astronomy, The University of Manchester, Manchester M13 9PL, UK\\
	$^{2}$ASTRON, the Netherlands Institute for Radio Astronomy, Postbus 2, NL-7990 AA Dwingeloo, The Netherlands 
}

\date{Accepted \today. Received \today; in original form \today}

\pubyear{2017}

\begin{document}
\label{firstpage}
\pagerange{\pageref{firstpage}--\pageref{lastpage}}
\maketitle

\begin{abstract}
We investigate the impact of intrinsic, kinematic and gravitational effects on high precision pulsar timing. We present an analytical derivation and a numerical computation of the impact of these effects on the first and second derivative of the pulsar spin frequency. In addition, in the presence of white noise, we derive an expression to determine the expected measurement uncertainty of a second derivative of the spin frequency for a given timing precision, observing cadence and timing baseline and find that it strongly depends on the latter ($\propto t^{-7/2}$). We show that for pulsars with significant proper motion, the spin frequency second derivative is dominated by a term dependent on the radial velocity of the pulsar. Considering the data sets from three Pulsar Timing Arrays, we find that for PSR\,J0437$-$4715 a detectable spin frequency second derivative will be present if the absolute value of the radial velocity exceeds 33\,km\,s$^{-1}$. Similarly, at the current timing precision and cadence, continued timing observations of PSR\,J1909$-$3744 for about another eleven years, will allow the measurement of its frequency second derivative and determine the radial velocity with an accuracy better than 14\,km\,s$^{-1}$. With the ever increasing timing precision and observing baselines, the impact of the, largely unknown, radial velocities of pulsars on high precision pulsar timing can not be neglected.

\end{abstract}


\section{Introduction}
Pulsars are good celestial clocks due to their rapid and stable rotation. Routinely measuring the times of arrival (ToAs) of the pulses from a pulsar, which is known as pulsar timing, helps to determine a series of important pulsar parameters (e.g. \citealt{ehm06}), describing the rotational, astrometric, and in some cases, the binary properties. Precise pulsar timing provides useful tools to probe fundamental physics, e.g. gravitational waves \citep{det79, hd83} and theories of gravity \citep{dt92,sta03,fwe12}. Currently, several dozen millisecond pulsars (MSPs), which form Pulsar Timing Arrays (PTAs) \citep{fb90}, are observed regularly to search for gravitational waves in the nano-Hz frequency band (e.g. \citealt{jll04,svc08,lwk11,dcl16,rhc16, abb15}).

To achieve these goals the highest possible timing precision is required and all other potentially measurable effects on the observed rate of pulsar spin need to be modeled. To achieve the former one can increase the time span of observations, or improve the sensitivity of the telescopes being used or building new telescopes, such as MeerKat (e.g. \citealt{kea17}), Five-hunderd-meter Aperture Spherical Telescope (FAST, e.g., \citealt{slk09, nlj11}) and the Square Kilometer Array (SKA, e.g., \citealt{sks09,laz13,kea17}). With this increased precision even more effects may become measurable. 

Some of these effects will be due to the motion of a pulsar and its location in the Galaxy. For the first derivative of spin frequency, $\dot{f}$, the pulsar proper motion (Shklovskii effect, \citealt{shk70}) and the Galactic acceleration (e.g. \citealt{dt91,nt95}) make noticeable contribtuions in may cases and their impacts have been well studied. The impact of kinematics and dynamics on the second derivative of the spin frequency, $\ddot{f}$, is less well studied. \cite{van03a} showed how the radial velocity of PSR J0437$-$4715 could contribute to the spin frequency second derivative, $\ddot{f}$, while \cite{phi92} investigated the dynamical influence of the ambient gravitational potential, such as that of a globular cluster, and its impact on the apparent $\ddot{f}$. Here we consider the intrinsic, kinematic and gravitational contributions to $\ddot{f}$ of MSPs with the aim of determining its overall amplitude and thus the effects on precision timing. Furthermore, we aim to find measurable quantities relating to the motion and the gravitational environment. The intrinsic component, $\ddot{f}_0$, is due to the spinning magnetic dipole model for pulsar rotation (e.g. \citealt{lk04}). For most MSPs, this component is small. However, a measurement of $\ddot{f}_0$ would constrain the braking index, and in turn the geometry and radiation properties (e.g. \citealt{bes99,ml99}) of an MSP.

One of the kinematic terms of interest is the radial velocity. Radial velocities are important in determining the full three dimensional motion of the pulsar through the Galaxy (see, e.g., \citealt{lwj09,bjs16}) thus useful in studying their formation and evolution (e.g. \citealt{fbw+11}). As has been done for globular clusters (e.g. \citealt{phi92,phi93,prf17} and \citealt{frk17}), the influence on $\ddot{f}$ of the Galactic potential means that a measurement may provide new tools to constrain the Galactic structure (e.g.\ \citealt{cl15}). The proper motion of MSPs is typically easily measured after just a couple of years of pulsar timing. However, because the radial velocity is inseparable from the observed spin frequency, it is not directly measurable. For pulsars in binary systems, the systemic radial velocity, i.e.\ that of the binary system as a whole, can be measured if the companion has suitable spectral lines to allow for orbital phase resolved spectroscopy.

The outline of this paper is as follows. We firstly present the expression of the frequency second derivative and introduce the different contributions in Section \ref{sec:derive}, then explain the pulsar sample and the numerical method used to compute the different contributions of $\ddot{f}$ in Section \ref{sec:galpy} and finally discuss its application to constrain the pulsar radial velocity and the braking index in Section \ref{sec:discuss}. Appendix \ref{apx:rdot} details the computation of the time derivatives of the position unit vector, Appendix \ref{apx:LSRtrans} the transformation of the velocities between the local standards of rest and Appendix \ref{apx:sigmaf2} the derivation of the measurement uncertainty of the frequency second derivative. 

\section{Frequency Derivatives}
\label{sec:derive}
The observed spin frequency of a pulsar with an intrinsic spin frequency $f_0$ located at position $\bmath{r}$ and moving with a velocity $\bmath{\varv}$ is subject to the Doppler effect through
\begin{equation}
\centering
f=\bigg(1-\frac{\bmath{\varv}\cdot\hat{\bmath{r}}}{c}\bigg)f_0,
\label{eqn:doppler}
\end{equation}
where $\hat{\bmath{r}}$ is the unit vector of $\bmath{r}$ and $c$ is the speed of light. Here, we have assumed that the velocity $\varv=|\bmath{\varv}|\ll c$. Differentiating Eqn.~\ref{eqn:doppler} with respect to time yields
\begin{equation}
\label{eqn:f1step1}
\dot{f}=\bigg(1-\frac{\bmath{\varv}\cdot\hat{\bmath{r}}}{c}\bigg)\dot{f}_0-\frac{\bmath{\varv}\cdot\dot{\hat{\bmath{r}}}}{c}f_0-\frac{\dot{\bmath{\varv}}\cdot\hat{\bmath{r}}}{c}f_0,
\end{equation}
and differentiating again yields 
\begin{equation}
\label{eqn:f2step1}
\ddot{f}=\bigg(1-\frac{\bmath{\varv}\cdot\hat{\bmath{r}}}{c}\bigg)\ddot{f}_0-\frac{2\bmath{\varv}\cdot\dot{\hat{\bmath{r}}}}{c}\dot{f}_0-\frac{2\dot{\bmath{\varv}}\cdot\hat{\bmath{r}}}{c}\dot{f}_0-\frac{\bmath{\varv}\cdot\ddot{\hat{\bmath{r}}}}{c}f_0-\frac{2\dot{\bmath{\varv}}\cdot\dot{\hat{\bmath{r}}}}{c}f_0-\frac{\ddot{\bmath{\varv}}\cdot\hat{\bmath{r}}}{c}f_0.
\end{equation}
Here, $\dot{f}_0$ and $\ddot{f}_0$ are the intrinsic first and second time derivatives of the spin frequency, $\dot{\bmath{\varv}}$ and $\ddot{\bmath{\varv}}$ are the first two time derivatives of the velocity vector, and $\dot{\hat{\bmath{r}}}$ and $\ddot{\hat{\bmath{r}}}$ are the first two time derivatives of the position unit vector.

To simplify Equations~\ref{eqn:doppler}, \ref{eqn:f1step1} and \ref{eqn:f2step1}, we denote the pulsar distance by $r=|\bmath{r}|$,  the acceleration by $\bmath{a}=\dot{\bmath{\varv}}$, the jerk by $\bmath{j}=\ddot{\bmath{\varv}}$ and decompose $\bmath{\varv}$, $\bmath{a}$ and $\bmath{j}$ into radial ($\parallel$) and transverse  ($\perp$) components. Using the relations in Appendix~\ref{apx:rdot}, we obtain
\begin{equation}
\label{eqn:fullf0}
f=\bigg(1-\frac{\varv_\parallel}{c}\bigg)f_0,
\end{equation}
\begin{equation} 
\label{eqn:fullf1}
\dot{f}=\bigg(1-\frac{\varv_\parallel}{c}\bigg)\dot{f}_0-\frac{\varv^2_\perp}{rc}f_0-\frac{a_\parallel}{c}f_0 \equiv \bigg(1-\frac{\varv_\parallel}{c}\bigg)\dot{f}_0 + \dot{f}_{\rm shk} + \dot{f}_{\rm acc},
\end{equation}
and 
\begin{equation}
\begin{split}
\label{eqn:fullf2}
\ddot{f}&=\bigg(1-\frac{\varv_\parallel}{c}\bigg)\ddot{f}_0-2\frac{\varv^2_\perp}{rc}\dot{f}_0-2\frac{a_\parallel}{c}\dot{f}_0+\frac{3\varv_\parallel \varv^2_\perp}{r^2 c}f_0-\frac{3\bmath{\varv}_\perp\cdot\bmath{a}_\perp}{rc}f_0-\frac{j_\parallel}{c}f_0 \\
 &\equiv\bigg(1-\frac{\varv_\parallel}{c}\bigg)\ddot{f}_0+\ddot{f}_{\rm shk} + \ddot{f}_{\rm acc}+\ddot{f}_\parallel + \ddot{f}_\perp + \ddot{f}_{\rm jerk},
 \end{split}
 \end{equation}
where $\equiv$ indicates definition and the meaning of each term is discussed in Section \ref{subsec:spindown} and \ref{subsec:acc}.

Eqn.~\ref{eqn:fullf1} is consistent with those derived by \cite{dt91} and \cite{phi92}, using the approximation that $f_0\approx f$. Expressions for $\ddot{f}$ were derived by \citet[eqn.~3.1]{phi92} and \citet[eqn.~3.28]{van03a} and are also  consistent with our derivation. We do however note that the numerical factor 2 in the third term of eqn.~3.1 by \cite{phi92} should be 1, and that the \cite{van03a} derivation does not consider acceleration and jerk. Furthermore, the \textsc{tempo} \citep{nds+15} and \textsc{tempo2} \citep{ehm06,hem06} software include an undocumented parameter to model a known radial velocity (N.\ Wex 1997, priv.\ comm.). 

Eqn.~\ref{eqn:fullf1} and~\ref{eqn:fullf2} are generic and depend on how the position, velocity and acceleration vectors are defined. Here, we are primarily interested in long timescale variations and hence define position and velocity between the Solar System Barycentre (SSB) and the position of an isolated pulsar or the binary barycentre in the case of a binary pulsar. This implicitly assumes that variations due to the motion of the observer around the SSB and the pulsar around the binary barycenter are either negligible or accounted for in the timing model. 
  
\subsection{Intrinsic spin-down}
\label{subsec:spindown}
Pulsars are believed to be well modeled as spinning magnetic dipoles and thus lose rotational kinetic energy through the emission of magnetic dipole radiation. Moreover they may also be affected by torques associated with the radio emission processes. This energy loss leads to a 
slowdown of the rotation of the pulsar, and hence the intrinsic spin frequency derivative $\dot{f}_0$. Depending on the configuration of the magnetic field, the spin down will give rise to higher order spin frequency derivatives. Under the assumption that the spin down follows a power law, the spin frequency second derivative can be written as $\ddot{f}_0=n\dot{f}_0^2/f_0$ where $n$ is the braking index.

A broad range of braking indices have been predicted by different radiation models. In the case of pure magnetic dipole radiation $n=3$, while alternative theories, including pulsar winds \citep{ml99,spi04,tko07}, current losses (e.g. \citealt{mal17}), evolution of the strength of magnetic field (e.g. \citealt{ho15}) and change in the inclination angle between the magnetic axis and the spin axis \citep{tko01,jk17}, allow $n$ to vary roughly from $-1$ to $5$ \citep{ho15,jk17}. Here the term braking process will be used to refer to all these mechanisms. For simplicity, in this paper we assume $n=3$, which is adequate for our order-of-magnitude estimates of $\ddot{f}_0$. $\ddot{f}_0$ is thus always positive as we use a $n>0$.

The first term of Eqn.~\ref{eqn:fullf0}, ~\ref{eqn:fullf1} and~\ref{eqn:fullf2} shows the modulations of the Doppler effect on the intrinsic value of $f_0$, $\dot{f}_0$ and $\ddot{f}_0$. Since the typical velocity of millisecond pulsar is several orders of magnitude smaller than the speed of light, we 
can safely ignore the factor $(1-\varv_\parallel/c)$. Using the observed value of $\dot{f}$ and accounting for the kinematic effects (the $\varv_\perp$ and $a_\parallel$ term) in Eqn.~\ref{eqn:fullf1}, the value of $\dot{f}_0$ is available and $\ddot{f}_0$ can thus be obtained through $\ddot{f}_0=n\dot{f}_0^2/f_0$.

\subsection{Kinematic corrections}
\label{subsec:acc}
The motion of the pulsar introduces several effects into the observed spin frequency derivatives, of which the Shklovskii effect, giving rise to $\dot{f}_{\rm shk}/f_0=-\varv^2_\perp/(rc)$, is well known. The Shklovskii effect also impacts $\ddot{f}$ through the intrinsic spin frequency derivative, leading to the second term in Eqn.~\ref{eqn:fullf2}, $\ddot{f}_{\rm shk}/\dot{f}_0=-2\varv^2_\perp/(rc)$. If the proper motion $\mu$ and distance $r$ to a pulsar are known, the transverse velocity $\varv_\perp=\mu r$ and thus $\dot{f}_{\rm shk}/f_0$ and $\ddot{f}_{\rm shk}/\dot{f}_0$ can be computed.

The Shklovskii effect introduces two further contributions to the second derivative of the spin frequency. The first, term 4 in Eqn.~\ref{eqn:fullf2}, is $\ddot{f}_\parallel/f_0=3\varv_\parallel\varv^2_\perp/(r^2c)$, and depends linearly on the radial velocity $\varv_\parallel$. Interestingly, when using proper motion instead of transverse velocity, the term no longer depends on distance; $\ddot{f}_\parallel/f_0=3\varv_\parallel\mu^2/c$. As $\mu^2$ is always non-negative, $\ddot{f}_\parallel$ takes on the sign of the radial velocity, and can be both positive and negative. The second contribution, term 5 in Eqn.~\ref{eqn:fullf2}, is $\ddot{f}_\perp/f_0=-3\bmath{\varv}_\perp\cdot\bmath{a}_\perp/(rc)$, and only depends on the transverse components of velocity $\bmath{\varv}$ and acceleration $\bmath{a}$.

As well as the contribution of the velocity to $\ddot{f}$ we also need to consider the impact of acceleration and jerk terms, which give rise to $\dot{f}_{\rm acc}/f_0=-a_\parallel/c$ (term 3 in Eqn.~\ref{eqn:fullf1}), $\ddot{f}_{\rm acc}/\dot{f}_0=-2a_\parallel/c$ (term 3 in Eqn.~\ref{eqn:fullf2}) and $\ddot{f}_{\rm jerk}/f_0=-j_\parallel/c$ (term 6 in Eqn.~\ref{eqn:fullf2}). Both acceleration and jerk depend on the gravitational potential in which the pulsar is situated. All the pulsars in our Galaxy are affected by the Galactic potential, while the motion of pulsars in globular clusters are also perturbed by the more dominant cluster potential \citep{phi92,phi93,prf17,frk17}. Furthermore, pulsars in wide binary systems experience a long-term secular influence from the companions (e.g. \citealt{jr97}, \citealt{bjs16} and \citealt{kkn16}). Since the effects due to globular clusters and distant pulsar companions are highly dependent on the particular systems, it is difficult to accurately estimate the contributions generally. In the remainder of this paper we focus on the impact of the Galactic potential. 

The Galactic acceleration is usually decomposed into the parts parallel to and perpendicular to the Galactic plane \citep{dt91,nt95}, by using a Galactic rotation curve and the empirical $K_z$ force law (i.e. the acceleration perpendicular to the Galactic plane, see \citealt{hf04}). In the computation of acceleration and jerk, using a spherically symmetric Galactic potential \citep{phi92} can not capture the influence of the complicated Galactic structure (e.g. \citealt{rmb14}), while using the rotation curve and the $K_z$ force law \citep{hf04} does not lead to the value of the jerk term. Here we combine the planar and vertical terms and compute $\bmath{a}$ and $\bmath{j}$ numerically by using a more realistic Galactic potential model as described in Sec.~\ref{sec:galpy}.

\section{Computing frequency derivatives}
\label{sec:galpy}
Computing the expected $\ddot{f}$ for a given pulsar consists of three steps. Firstly, a Galactic potential model is required to compute the dynamical quantities including $\bmath{a}$ and $\bmath{j}$. Secondly, $\dot{f}_0$ is obtained through Eqn.~\ref{eqn:fullf1} by using the observables $r$, $\varv_\perp=\mu r$, $f$ and $\dot{f}$. Finally, each component in $\ddot{f}$ is found by using $\dot{f}_0$ and by assuming a radial velocity.

\subsection{Pulsars and Astrometric data}
\label{sec:pulsardata}
In this paper, we concentrate on MSPs observed by the PTAs. Rotational ($f$, $\dot{f}$) and astrometric ($\alpha$, $\delta$, $\mu_\alpha$, $\mu_\delta$, $\pi$) parameters of MSPs timed by the European Pulsar Timing Array (EPTA), Parkes Pulsar Timing Array (PPTA) and the North American Nanoherts Observatory for Gravitational Waves (NANOGrav) are provided by \cite{dcl16} (EPTA), \cite{rhc16} (PPTA) and \cite{abb15,mnf16} (NANOGrav), respectively. In the case of NANOGrav data, rotational parameters are taken from \cite{abb15} and astrometric parameters in equatorial coordinates from \cite{mnf16}. Since several pulsars are observed by more than one PTA, we use data from all 42 pulsars timed by the EPTA, 6 out of 20 pulsars timed by the PPTA but not the EPTA, and 14 out of 37 NANOgrav pulsars not timed by either EPTA or PPTA. The total sample of 62 pulsars is given in Table~\ref{tab:ajf}. This list does not include PSRs J0931$-$1902, B1821$-$24A and J1832$-$0836. Though these pulsars are timed by PPTA or NANOgrav, no significant proper motion has been measured for these pulsars. We note however that because PSR B1821$-$24A is located in a globular cluster it is likely that
the timing will need to take into account an $\ddot{f}$ contribution from the jerk due to the globular cluster potential.

\begin{table*}
\label{tab:ajf}
\scriptsize
\centering
\begin{tabular}{lrcrrrrrrrrrr}
\hline 
\hline
 
 \multicolumn{1}{c}{PSR} &\multicolumn{1}{c}{$f$} & \multicolumn{1}{c}{$r$} & \multicolumn{1}{c}{$r$} & \multicolumn{1}{c}{$\mu$} & \multicolumn{1}{c}{$\varv_{\rm LSR}$} & \multicolumn{1}{c}{$\dot{f}_0$} &\multicolumn{1}{c}{ $a_\parallel/c$} & \multicolumn{1}{c}{$\varv_\perp^2/(rc)$}  & \multicolumn{1}{c}{$3\varv_\perp^2/(r^2c)$}  & \multicolumn{1}{c}{$j_\parallel/c$ slope} & \multicolumn{1}{c}{$j_\parallel/c$ incpt}  & \multicolumn{1}{c}{$3\boldsymbol{\varv}_\perp\cdot\boldsymbol{a}_\perp/(rc)$}  \\ [1pt] 
      &   &  & & &  & \multicolumn{1}{c}{($10^{-16}$)} & \multicolumn{1}{c}{($10^{-19}$)}  & \multicolumn{1}{c}{($10^{-19}$)}  & \multicolumn{1}{c}{($10^{-34}$)} & \multicolumn{1}{c}{($10^{-36}$)} & \multicolumn{1}{c}{($10^{-34}$)}  & \multicolumn{1}{c}{($10^{-34}$)}  \\  
& \multicolumn{1}{c}{(Hz)}  & \multicolumn{1}{c}{(kpc)} &\multicolumn{1}{c}{type} &\multicolumn{1}{c}{(mas yr$^{-1}$)} & \multicolumn{1}{c}{(km s$^{-1}$)} & \multicolumn{1}{c}{(s$^{-2}$)} & \multicolumn{1}{c}{(s$^{-1}$)} & \multicolumn{1}{c}{(s$^{-1}$)} & \multicolumn{1}{c}{(km$^{-1}$s$^{-1}$)} & \multicolumn{1}{c}{(km$^{-1}$s$^{-1})$} & \multicolumn{1}{c}{(s$^{-2})$}  & \multicolumn{1}{c}{(s$^{-2})$} \\	
       \hline	
J0023$+$0923$^\dagger$    & 328 & 0.70   &  DM  &  14.01   & $  -1.25   $ & $ -11.73  $ & $ -1.66  $ &   3.34    &  0.46    & $  -2.52   $ & $  -1.82  $ & $  -4.92  $ \\
J0030$+$0451$^\ast$      & 206 & 0.35   &  PX  &  5.90    & $  1.36    $ & $ -4.51   $ & $ -1.32  $ &   0.30    &  0.08    & $  -6.74   $ & $  -0.75  $ & $  -0.30  $ \\
J0034$-$0534$^\ast$      & 533 & 0.54   &  DM  &  12.13   & $  3.24    $ & $ -14.07  $ & $ -1.81  $ &   1.93    &  0.35    & $  -3.87   $ & $  -1.00  $ & $  -2.38  $ \\
J0218$+$4232$^\ast$      & 430 & 3.15   &  PX  &  6.17    & $  -31.20  $ & $ -141.88 $ & $ 0.67   $ &   2.92    &  0.09    & $  1.20    $ & $  4.37   $ & $  5.34   $ \\
J0340$+$4130$^\dagger$    & 303 & 1.70   &  DM  &  5.17    & $  -10.20  $ & $ -5.91   $ & $ 0.86   $ &   1.10    &  0.06    & $  1.80    $ & $  1.52   $ & $  -1.86  $ \\
J0437$-$4715$^\flat$      & 174 & 0.16   &  PX  &  140.91  & $  16.60   $ & $ -4.23   $ & $ -0.51  $ &   75.68   &  46.90   & $  -9.77   $ & $  6.91   $ & $  23.87  $ \\
J0610$-$2100$^\ast$      & 259 & 3.54   &  DM  &  19.04   & $  57.50   $ & $ -0.10   $ & $ 0.29   $ &   31.18   &  0.86    & $  0.93    $ & $  -1.05  $ & $  25.13  $ \\
J0613$-$0200$^\ast$      & 327 & 0.78   &  PX  &  10.51   & $  25.60   $ & $ -9.44   $ & $ 0.32   $ &   2.09    &  0.26    & $  1.49    $ & $  -2.22  $ & $  -3.83  $ \\
J0621$+$1002$^\ast$      &  35 & 1.36   &  DM  &  3.27    & $  26.20   $ & $ -0.52   $ & $ 1.01   $ &   0.35    &  0.03    & $  2.15    $ & $  -1.25  $ & $  -0.30  $ \\
J0645$+$5158$^\dagger$    & 113 & 0.80   &  PX  &  7.63    & $  -0.64   $ & $ -0.47   $ & $ 0.30   $ &   1.13    &  0.14    & $  2.05    $ & $  1.48   $ & $  1.84   $ \\
J0711$-$6830$^\flat$       & 182 & 0.85   &  DM  &  21.10   & $  9.80    $ & $ -3.50   $ & $ -1.31  $ &   9.24    &  1.05    & $  -3.86   $ & $  -1.22  $ & $  -7.44  $ \\
J0751$+$1807$^\ast$      & 287 & 1.07   &  PX  &  13.68   & $  20.00   $ & $ -4.94   $ & $ 0.34   $ &   4.86    &  0.44    & $  1.95    $ & $  -0.93  $ & $  1.55   $ \\
J0900$-$3144$^\ast$      &  90 & 0.81   &  PX  &  2.26    & $  19.30   $ & $ -4.01   $ & $ -0.68  $ &   0.10    &  0.01    & $  -2.34   $ & $  -0.70  $ & $  0.29   $ \\
J1012$+$5307$^\ast$      & 190 & 1.15   &  PX  &  25.62   & $  -5.49   $ & $ -2.86   $ & $ -0.76  $ &   18.30   &  1.55    & $  1.04    $ & $  1.36   $ & $  -5.83  $ \\
J1022$+$1001$^\ast$      &  61 & 1.09   &  PX  &  18.45   & $  10.70   $ & $ -1.13   $ & $ -1.29  $ &   9.03    &  0.80    & $  -0.33   $ & $  1.93   $ & $  12.66  $ \\
J1024$-$0719$^\ast$      & 194 & 1.08   &  PX  &  59.72   & $  12.70   $ & $ 10.93   $ & $ -1.45  $ &   93.82   &  8.42    & $  -1.47   $ & $  -0.34  $ & $  26.29  $ \\
J1045$-$4509$^\flat$      & 134 & 0.34   &  PX  &  7.99    & $  6.61    $ & $ -3.14   $ & $ -0.37  $ &   0.53    &  0.15    & $  -3.57   $ & $  -0.01  $ & $  -0.67  $ \\
J1455$-$3330$^\ast$      & 125 & 0.80   &  PX  &  8.19    & $  -15.10  $ & $ -3.68   $ & $ -0.31  $ &   1.30    &  0.16    & $  -0.88   $ & $  2.99   $ & $  6.23   $ \\
J1600$-$3053$^\ast$      & 278 & 1.49   &  PX  &  7.00    & $  -21.40  $ & $ -6.70   $ & $ 0.53   $ &   1.78    &  0.12    & $  1.27    $ & $  2.65   $ & $  3.42   $ \\
J1603$-$7202$^\flat$      &  67 & 0.53   &  PX  &  7.73    & $  -5.17   $ & $ -0.67   $ & $ -0.19  $ &   0.77    &  0.14    & $  -1.38   $ & $  -0.08  $ & $  -3.17  $ \\
J1614$-$2230$^\dagger$    & 317 & 0.65   &  PX  &  32.22   & $  -13.60  $ & $ -4.46   $ & $ 0.10   $ &   16.40   &  2.45    & $  0.85    $ & $  4.78   $ & $  15.58  $ \\
J1640$+$2224$^\ast$      & 316 & 1.16   &  DM  &  11.48   & $  -6.68   $ & $ -2.14   $ & $ -1.57  $ &   3.71    &  0.31    & $  -2.87   $ & $  1.04   $ & $  7.71   $ \\
J1643$-$1224$^\ast$      & 216 & 0.76   &  PX  &  7.28    & $  -12.00  $ & $ -8.41   $ & $ 0.09   $ &   0.98    &  0.13    & $  0.80    $ & $  0.26   $ & $  1.59   $ \\
J1713$+$0747$^\ast$      & 219 & 1.11   &  PX  &  6.29    & $  -5.16   $ & $ -3.97   $ & $ -0.55  $ &   1.06    &  0.09    & $  -1.10   $ & $  0.29   $ & $  6.68   $ \\
J1721$-$2457$^\ast$      & 286 & 1.30   &  DM  &  25.07   & $  -11.70  $ & $ 1.54    $ & $ 1.42   $ &   19.85   &  1.48    & $  4.00    $ & $  4.49   $ & $  11.21  $ \\
J1730$-$2304$^\ast$      & 123 & 0.90   &  PX  &  22.57   & $  -10.90  $ & $ -1.56   $ & $ 0.97   $ &   11.19   &  1.20    & $  3.79    $ & $  1.41   $ & $  4.21   $ \\
J1732$-$5049$^\flat$      & 188 & 1.39   &  DM  &  9.88    & $  -19.60  $ & $ -4.24   $ & $ 0.89   $ &   3.30    &  0.23    & $  2.07    $ & $  -1.02  $ & $  -8.91  $ \\
J1738$+$0333$^\ast$      & 171 & 1.45   &  PX  &  8.65    & $  0.77    $ & $ -6.60   $ & $ -0.05  $ &   2.64    &  0.18    & $  -0.21   $ & $  -2.78  $ & $  0.17   $ \\
J1741$+$1351$^\dagger$    & 267 & 0.90   &  DM  &  11.60   & $  -6.36   $ & $ -20.87  $ & $ -0.51  $ &   2.94    &  0.32    & $  -1.55   $ & $  -0.94  $ & $  -0.39  $ \\
J1744$-$1134$^\ast$      & 245 & 0.42   &  PX  &  21.01   & $  -11.80  $ & $ -4.20   $ & $ 0.32   $ &   4.49    &  1.04    & $  2.51    $ & $  1.38   $ & $  5.12   $ \\
J1747$-$4036$^\dagger$    & 608 & 3.30   &  DM  &  6.00    & $  -33.30  $ & $ -44.44  $ & $ 3.91   $ &   2.89    &  0.09    & $  4.35    $ & $  -0.79  $ & $  -12.59 $ \\
J1751$-$2857$^\ast$      & 255 & 1.10   &  DM  &  8.56    & $  -10.70  $ & $ -6.47   $ & $ 1.35   $ &   1.96    &  0.17    & $  4.55    $ & $  0.29   $ & $  0.48   $ \\
J1801$-$1417$^\ast$      & 276 & 1.52   &  DM  &  11.30   & $  -1.58   $ & $ -2.30   $ & $ 1.57   $ &   4.71    &  0.30    & $  3.75    $ & $  -2.57  $ & $  -1.01  $ \\
J1802$-$2124$^\ast$      &  79 & 0.64   &  PX  &  3.21    & $  -9.97   $ & $ -4.49   $ & $ 0.71   $ &   0.16    &  0.02    & $  3.87    $ & $  -0.24  $ & $  0.32   $ \\
J1804$-$2717$^\ast$      & 107 & 0.78   &  DM  &  17.19   & $  -9.96   $ & $ -3.98   $ & $ 0.90   $ &   5.60    &  0.70    & $  4.08    $ & $  -0.52  $ & $  -0.95  $ \\
J1843$-$1113$^\ast$      & 542 & 1.09   &  PX  &  3.73    & $  -2.19   $ & $ -27.35  $ & $ 0.90   $ &   0.37    &  0.03    & $  2.77    $ & $  -1.09  $ & $  1.61   $ \\
J1853$+$1303$^\ast$      & 244 & 0.88   &  PX  &  3.22    & $  -4.13   $ & $ -5.17   $ & $ -0.05  $ &   0.22    &  0.02    & $  -0.55   $ & $  -1.37  $ & $  1.31   $ \\
B1855$+$09$^\ast$        & 186 & 1.10   &  PX  &  6.02    & $  -0.29   $ & $ -6.01   $ & $ 0.07   $ &   0.97    &  0.09    & $  -0.20   $ & $  -1.19  $ & $  3.33   $ \\
J1903$+$0327$^\dagger$    & 465 & 6.00   &  DM  &  7.04    & $  76.50   $ & $ -39.38  $ & $ -4.39  $ &   7.22    &  0.12    & $  -7.92   $ & $  -8.20  $ & $  25.86  $ \\
J1909$-$3744$^\ast$      & 339 & 1.15   &  PX  &  37.02   & $  -8.13   $ & $ -3.11   $ & $ 0.29   $ &   38.15   &  3.24    & $  1.28    $ & $  -1.19  $ & $  -3.95  $ \\
J1910$+$1256$^\ast$      & 201 & 0.55   &  PX  &  7.38    & $  -8.64   $ & $ -3.75   $ & $ -0.01  $ &   0.73    &  0.13    & $  -0.27   $ & $  -0.45  $ & $  1.91   $ \\
J1911$-$1114$^\ast$      & 276 & 1.23   &  DM  &  16.49   & $  1.12    $ & $ -8.21   $ & $ 0.59   $ &   8.12    &  0.64    & $  1.40    $ & $  3.11   $ & $  13.98  $ \\
J1911$+$1347$^\ast$      & 216 & 2.07   &  DM  &  4.73    & $  13.90   $ & $ -7.79   $ & $ -0.55  $ &   1.13    &  0.05    & $  -1.94   $ & $  -2.95  $ & $  4.66   $ \\
J1918$-$0642$^\ast$      & 131 & 1.24   &  DM  &  9.31    & $  2.40    $ & $ -4.00   $ & $ 0.41   $ &   2.61    &  0.20    & $  0.77    $ & $  0.48   $ & $  7.67   $ \\
J1923$+$2515$^\dagger$    & 264 & 1.60   &  DM  &  15.93   & $  2.00    $ & $ -4.32   $ & $ -1.00  $ &   9.87    &  0.60    & $  -2.73   $ & $  0.55   $ & $  10.12  $ \\
B1937$+$21$^\ast$        & 642 & 3.27   &  PX  &  0.41    & $  17.80   $ & $ -434.59 $ & $ -2.49  $ &   0.01    &  0.00    & $  -3.83   $ & $  -4.00  $ & $  0.35   $ \\
J1944$+$0907$^\dagger$    & 193 & 1.80   &  DM  &  27.20   & $  10.90   $ & $ -0.33   $ & $ -0.61  $ &   32.36   &  1.75    & $  -2.02   $ & $  -5.74  $ & $  -7.42  $ \\
J1949$+$3106$^\dagger$    &  76 & 3.60   &  DM  &  16.40   & $  3.16    $ & $ -4.14   $ & $ -3.32  $ &   23.52   &  0.64    & $  -3.46   $ & $  0.08   $ & $  -9.09  $ \\
B1953$+$29$^\ast$        & 163 & 4.64   &  DM  &  4.76    & $  2.20    $ & $ -8.21   $ & $ -4.36  $ &   2.56    &  0.05    & $  -3.26   $ & $  -1.10  $ & $  2.81   $ \\
J2010$-$1323$^\ast$      & 191 & 1.02   &  DM  &  6.24    & $  -0.00   $ & $ -1.69   $ & $ -0.52  $ &   0.96    &  0.09    & $  -1.11   $ & $  -2.28  $ & $  -3.23  $ \\
J2017$+$0603$^\dagger$    & 345 & 1.60   &  DM  &  2.26    & $  7.75    $ & $ -9.85   $ & $ -1.10  $ &   0.20    &  0.01    & $  -2.56   $ & $  -3.58  $ & $  -2.73  $ \\
J2019$+$2425$^\ast$      & 254 & 1.50   &  DM  &  21.76   & $  -0.56   $ & $ -0.41   $ & $ -1.20  $ &   17.25   &  1.12    & $  -3.10   $ & $  0.25   $ & $  7.91   $ \\
J2033$+$1734$^\ast$      & 168 & 2.00   &  DM  &  10.85   & $  6.30    $ & $ -2.48   $ & $ -1.86  $ &   5.71    &  0.28    & $  -3.39   $ & $  -0.96  $ & $  6.34   $ \\
J2043$+$1711$^\dagger$    & 420 & 1.30   &  PX  &  12.37   & $  -0.19   $ & $ -7.76   $ & $ -1.28  $ &   4.83    &  0.36    & $  -3.26   $ & $  -0.64  $ & $  3.80   $ \\
J2124$-$3358$^\ast$      & 203 & 0.38   &  PX  &  52.07   & $  -3.07   $ & $ -3.55   $ & $ -0.97  $ &   25.16   &  6.40    & $  -5.62   $ & $  1.27   $ & $  7.33   $ \\
J2129$-$5721$^\flat$       & 268 & 1.69   &  DM  &  13.32   & $  -9.22   $ & $ -13.65  $ & $ -2.17  $ &   7.26    &  0.42    & $  -2.76   $ & $  -2.20  $ & $  -10.69 $ \\
J2145$-$0750$^\ast$      &  62 & 0.65   &  PX  &  13.05   & $  -2.25   $ & $ -1.08   $ & $ -1.44  $ &   2.67    &  0.40    & $  -4.25   $ & $  -0.27  $ & $  4.70   $ \\
J2214$+$3000$^\dagger$    & 321 & 1.50   &  DM  &  20.07   & $  -10.30  $ & $ -11.04  $ & $ -1.88  $ &   14.68   &  0.95    & $  -2.78   $ & $  1.17   $ & $  -6.98  $ \\
J2229$+$2643$^\ast$      & 336 & 1.43   &  DM  &  6.07    & $  -9.36   $ & $ -1.93   $ & $ -1.93  $ &   1.28    &  0.09    & $  -2.70   $ & $  -0.40  $ & $  -3.14  $ \\
J2302$+$4442$^\dagger$    & 193 & 1.10   &  DM  &  3.61    & $  -15.40  $ & $ -5.24   $ & $ -1.03  $ &   0.35    &  0.03    & $  -2.26   $ & $  0.43   $ & $  -1.82  $ \\
J2317$+$1439$^\ast$      & 290 & 1.01   &  PX  &  3.54    & $  -5.04   $ & $ -2.52   $ & $ -1.91  $ &   0.31    &  0.03    & $  -2.72   $ & $  -0.27  $ & $  1.97   $ \\
J2322$+$2057$^\ast$      & 208 & 0.80   &  DM  &  23.99   & $  -6.43   $ & $ -2.17   $ & $ -1.55  $ &   11.19   &  1.36    & $  -3.13   $ & $  -2.09  $ & $  -6.97  $ \\
	\hline                                    
  \end{tabular}                             
  \caption{The pulsar parameters: columns are pulsar name, apparent spin frequency, distance and the distance type (inferred from DM or parallax), the proper motion, the predicted radial LSR velocity, the intrinsic spin frequency derivative, the radial acceleration term, two transverse velocity terms, the slope and intercept of radial jerk, the mixing term of transverse velocity and transverse acceleration. The superscripts in the pulsar names in the first column indicate the PTAs ($\ast$ for EPTA, $\flat$ for PPTA and $\dagger$ for NANOGrav) where the astrometric pulsar data came from. Numbers in brackets in the second row of the header are the orders of magnitude.}
  \label{tab:ajf}
\end{table*}

Where available, we use parallax distances that have been corrected for the Lutz-Kelker bias \citep{lk73} as provided by \cite{dcl16,rhc16} and \cite{mnf16}. For those pulsars for which a parallax distance is not available, we use the distance inferred from the observed dispersion measure and the NE2001 model of electron distribution \citep{cl01}.

The remaining parameter required to compute the $\ddot{f}$ contributions is the radial velocity $\varv_\parallel$. This parameter is unknown for all but the five pulsars listed in Table~\ref{tab:psrv} where the radial velocity is measured through optical spectroscopy of their binary companions. To compute $\ddot{f}_\parallel$ and the motion of the pulsar in the Galaxy, we vary the radial velocity from $-200$ to 200\,km\,s$^{-1}$, enough to cover the likely radial velocity of most MSPs. We note that $\varv_\parallel$ is the radial velocity as observed at SSB. It will therefore be a combination of an intrinsic radial velocity and the radial velocity of the local standard of rest (LSR) of the pulsar. We find that for the pulsars studied here, the radial component of the pulsar LSR velocity is $|\varv_{\rm LSR}|<80$\,km\,s$^{-1}$ (see Appendix~\ref{apx:LSRtrans} and Table~\ref{tab:ajf}).

\begin{table}
\caption{The frequency, radial velocity, proper motion and the theoretical $\ddot{f}_\parallel$ of the five pulsars with measured radial velocities. The references are: (1) \protect\cite{cgk98}, (2) \protect\cite{bjs16}, (3) \protect\cite{avk+12}, (4) 
   \protect\cite{fbw+11}, (5) \protect\cite{ksf+12}, (6) \protect\cite{ant13}, (7) \protect\cite{dcl16}, (8) \protect\cite{mnf16}}
   \label{tab:psrv}
  \begin{tabular}{lcccc}
   \hline\hline
   \multicolumn{1}{l}{PSR}& \multicolumn{1}{c}{$f$} 
   & \multicolumn{1}{c}{$\varv_\parallel$} 
   &\multicolumn{1}{c}{$\mu$} 
   &\multicolumn{1}{c}{$\ddot{f}_\parallel$}\\
    & \multicolumn{1}{c}{(Hz)} 
   & \multicolumn{1}{c}{(km s$^{-1}$)} 
   &\multicolumn{1}{c}{(mas yr$^{-1}$)} &\multicolumn{1}{c}{($10^{-30}$\,s$^{-3}$)}\\
   \hline
   J1012+5307 & 190 & $44\pm8^{(1)}$ & $25.6^{(7)}$ & $1.3\pm0.2$\\
   J1024$-$0719 & 194 & $185\pm4^{(2)}$ & $59.7^{(7)}$ & $30.2\pm0.7$ \\
   J1738+0333 & 171 & $-42\pm16^{(3)}$ & $8.65^{(7)}$ &$-0.13\pm0.05$ \\
   J1903+0327 & 465 & $42.1\pm2.5^{(4, 5)}$ & $7.04^{(8)}$ & $0.23\pm0.07$ \\
   J1909$-$3744 & 339 &$-73\pm30^{(6)}$ & $37.0^{(7)}$ & $-8\pm3$ \\
   \hline
   \end{tabular}
\end{table}

\subsection{Numerically computing Galactic dynamics}
We use the \texttt{MWPotential2014} Galactic potential and the orbit integrator implemented in the \textsc{GALPY} software package by \cite{bov15} to numerically integrate the motion of each pulsar, using their known right ascension, declination, distance, proper motion and the radial velocity. The motion of the pulsar through the Galaxy, expressed in Cartesian Galactic coordinates, was calculated over a time range from $-5$ to $+5$\,kyr. Each of these coordinates was fitted with a third degree polynomial to model the motion as a function of time and the polynomial coefficients were used to obtain the acceleration $\bmath{a}$ and jerk $\bmath{j}$, besides the position $\bmath{r}$ and velocity $\bmath{\varv}$ at time $t=0$\,kyr. Similar polynomials were fitted to the motion of the Sun over the same time range to obtain the relative acceleration and jerk between each pulsar and the Sun. Throughout our calculations we assumed a Galactocentric radius of $R_0=8.34$\,kpc and a circular rotation speed of $\varv_0=240$\,km\,s$^{-1}$ \citep{rmb14}, with the solar motion values by \cite{hbr+05}.

\subsection{The impact of field stars on \texorpdfstring{$\ddot{f}$}{f2}}
Another possible contribution to the frequency second derivative can come from field stars, which may affect the pulsar gravitationally. The first impact is the gravitational jerk (the $\ddot{f}_{\rm jerk}$ term), which can be approximated by $\ddot{f}_{\rm jerk}/f_0=-G\rho(\bmath{\varv}_{\rm s}-\bmath{\varv}_{\rm p})\cdot\hat{\bmath{r}}_{\rm p}/c$ \citep[section 3]{phi92}, where $G$ is the gravitational constant, $\rho$ is the local stellar mass density around the pulsar, and $\bmath{\varv}_{\rm s}$ and $\bmath{\varv}_{\rm p}$ are the velocities of the field star and the pulsar, respectively. Using a Galactic mid-plane stellar density of $\rho=0.04$ M$_\odot$\,pc$^{-3}$ \citep{bov17} and assuming a spin frequency of 1000 Hz and $(\bmath{\varv}_{\rm s}-\bmath{\varv}_{\rm p})\cdot\hat{\bmath{r}}_{\rm p}=100$\,km\,s$^{-1}$, we find $\ddot{f}_{\rm jerk}=-6\times10^{-32}$\,s$^{-3}$. The second impact is $\ddot{f}_{\rm acc}$ and can be estimated by $\ddot{f}_{\rm acc}/\dot{f}_0=-2G\rho(\bmath{r}_{\rm s}-\bmath{r}_{\rm p})\cdot\bmath{r}_{\rm p}/(r_{\rm p}c)$, where $\bmath{r}_{\rm s}$ and $\bmath{r}_{\rm p}$ are the position vectors of the star and the pulsar, respectively. Assuming a distance of 1 kpc for $(\bmath{r}_{\rm s}-\bmath{r}_{\rm p})\cdot\bmath{r}_{\rm p}/r_{\rm p}$ and a typical value of $-1\times10^{-16}$ s$^{-2}$ for $\dot{f}_0$, then $\ddot{f}_{\rm acc}=4\times10^{-36}$ s$^{-3}$. The third influence is $\ddot{f}_\perp/f_0=-3\bmath{\varv}_\perp\cdot\bmath{a}_\perp/(r_{\rm p}c)$. An upper limit can thus be set as $|\ddot{f}_\perp/f_0|\leq3\varv_\perp a/(r_{\rm p}c)=3\varv_\perp G\rho|\bmath{r}_{\rm s}-\bmath{r}_{\rm p}|/(r_{\rm p}c)$. Assuming $\varv_\perp=100$ km s$^{-1}$ and $r_{\rm p}=1$ kpc, then $|\ddot{f}_\perp|\leq2\times10^{-31}$\,s$^{-3}$. Considering the stringent requirement on the orientation of $\bmath{\varv}_\perp$ and $\bmath{a}_\perp$ to maximize the value of $\bmath{\varv}_\perp\cdot\bmath{a}_\perp$ in $\ddot{f}_\perp$, the expected value of $|\ddot{f}_\perp|$ should be much smaller than the nominal upper limit. Therefore, all three terms are generally much smaller than $2\times10^{-31}$ s$^{-3}$. 

\subsection{Magnitudes of \texorpdfstring{$\ddot{\MakeLowercase{f}}$}{f2} components}
Table~\ref{tab:ajf} gives the values of the important intermediate quantities. We list the values of $\dot{f}_0$, $a_\parallel/c$, $\varv^2_\perp/(rc)$ and $3\bmath{\varv}_\perp\cdot\bmath{a}_\perp/(rc)$ as these terms are independent of the radial velocity. $\ddot{f}_\parallel$ depends on the radial velocity and only its factor $3\varv_\perp^2/(r^2c)$ is listed as it can simply be multiplied by the radial velocity $\varv_\parallel$ to get its size. The jerk term $\ddot{f}_{\rm jerk}=-j_\parallel f_0/c$ has, to first order, a linear dependence on the radial velocity, and to allow comparison with the other effects, we provide the intercept (at $\varv_\parallel=0$\,km\,s$^{-1}$) and slope of $-j_\parallel/c$ with $\varv_\parallel$.

\begin{figure*}
 \includegraphics[width=18.5cm]{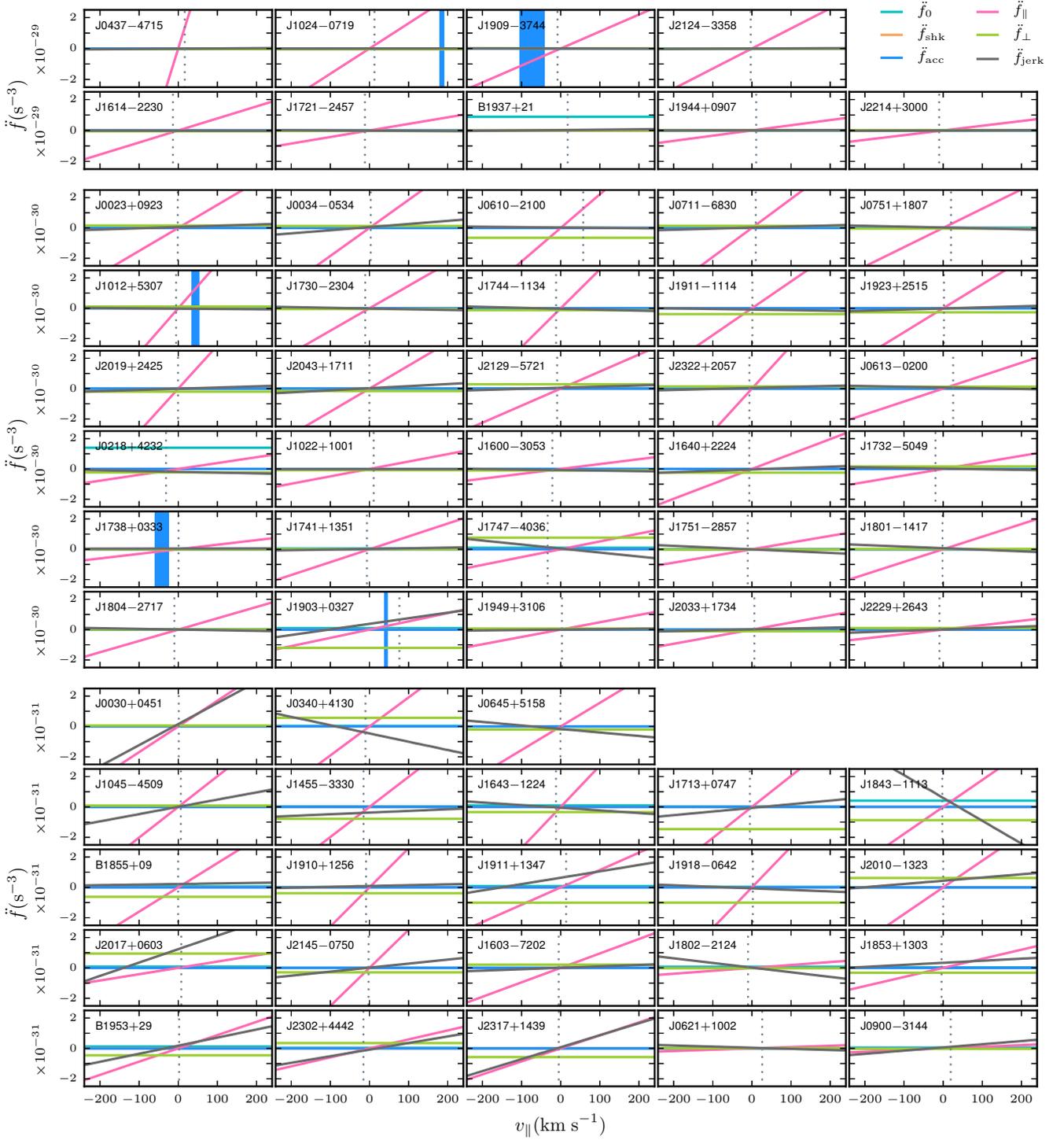}
 \caption{The six components in the frequency second derivative as a function of the radial velocity. The 62 pulsars are separated into three groups according to the order-of-magnitude of $\ddot{f}$. The vertical dashed lines indicate the predicted pulsar $\varv_{\rm LSR}$ in Table~\ref{tab:ajf}. The vertical blue shaded region marks the measured radial velocity and its uncertainty of the five pulsars in Table~\ref{tab:psrv}.}
 \label{fig:f2terms}
\end{figure*}

Using the intermediate quantities, the $\ddot{f}$ components in Eqn.~\ref{eqn:fullf2} can be computed. The impact of field stars is generally much smaller than the other contributions to $\ddot{f}$ and so are not included. Fig.~\ref{fig:f2terms} shows the value of these remaining components as a function of the radial velocity.

\subsection{Expected measurement uncertainty of \texorpdfstring{$\ddot{f}$}{f2}}
The timing ephemerides presented by \cite{dcl16,rhc16} and \cite{abb15} did not fit for the spin frequency second derivative $\ddot{f}$. Hence, measured $\ddot{f}$ values and their uncertainties are not available. To estimate the expected measurement uncertainty on $\ddot{f}$, given the other parameters in a timing ephemeris, we derive an analytical expression for $\sigma(\ddot{f})$ in Appendix~\ref{apx:sigmaf2}. Under the assumption that the ToA measurements of a pulsar with a spin frequency $f$ have equal, Gaussian root-mean-square (rms) uncertainties of $\sigma_{\rm rms}$ and are equally distributed at a cadence $\Delta t$ over a timespan $T$, the uncertainty $\sigma(\ddot{f})$ is
\begin{equation}
\sigma(\ddot{f})=2.3\times10^{-29}{\rm s}^{-3}\bigg(\frac{f}{500\ \mathrm{Hz}}\bigg)\bigg(\frac{\sigma_{\rm rms}}{1\ \mu{\rm s}}\bigg)\bigg(\frac{\Delta t}{10\ \rm d}\bigg)^{\frac{1}{2}}\bigg(\frac{T}{20\ \rm yr}\bigg)^{-\frac{7}{2}}.
\label{eqn:sigf2}
\end{equation}

To verify the validity of Eqn.~\ref{eqn:sigf2},  we used TEMPO2 \citep{hem06,ehm06} to simulate timing residuals and obtain the measurement uncertainty of $\ddot{f}$ by fitting the residuals with an ephemeris that includes $\ddot{f}$. We performed simulations over various spin frequencies ($f$ from 100 to 1000 Hz), rms timing residuals ($\sigma_{\rm rms}$ from 0.1 to 100 $\upmu$s), cadence ($\Delta t$ from 1 to 60 days) and timespan ($T$ from 10 to 50 years), which confirm both the parameter dependencies and the constant found in the analytical derivation of Eqn.~\ref{eqn:sigf2}. The relation in Eqn.~\ref{eqn:sigf2} is also consistent with that derived by \cite{bra87}.

Eqn.\,\ref{eqn:sigf2} shows that both the residual rms $\sigma_{\rm rms}$ and the timing span $T$ are dominant factors of the measurement uncertainty $\sigma(\ddot{f})$. The smaller the residual rms or the longer the timing span is, the lower the measurement uncertainty will be. In practice, the real improvement of $\sigma(\ddot{f})$ can be faster than the theoretical expectation, as with the timing span increasing, the residual rms can also be reduced due to the longer and probably better data sets. According to Eqn.\,\ref{eqn:sigf2}, $\sigma(\ddot{f})$ also depends on the cadence $\Delta t$ although the dependence is very weak. Since we did not specify the origin of $\ddot{f}$ in the derivation, Eqn.~\ref{eqn:sigf2} is applicable to estimate the measurement uncertainty of $\ddot{f}$ caused by any process. 

\section{Results and Discussion}
\label{sec:discuss}
The results shown in Fig.\,\ref{fig:f2terms}, and listed in
Table\,\ref{tab:ajf}, indicate that for the majority of the PTA MSPs,
the radial velocity induced contribution to the second derivative of
the spin frequency, $\ddot{f}_\parallel$, will dominate for relatively
large radial velocities ($|\varv_\parallel|\geqslant
50$\,km\,s$^{-1}$). For pulsars with high proper motions,
$\ddot{f}_\parallel$ can reach an order-of-magnitude of $10^{-29}$
s$^{-3}$. In addition, the impact of braking processes for pulsars with
low characteristic ages, like PSR\,B1937+21 and PSR J0218+4232, can
also be significant, with values of $10^{-29}$ and
$10^{-30}$\,s$^{-3}$, respectively for the case where $n=3$. The
contributions from the perpendicular velocity ($\ddot{f}_{\rm shk}$),
the Galactic acceleration ($\ddot{f}_{\rm acc}$) and jerk terms
($\ddot{f}_{\rm jerk}$) are generally less than $10^{-30}$ s$^{-3}$,
although $\ddot{f}_\perp$ and $\ddot{f}_{\rm jerk}$ are likely to be
comparable with $\ddot{f}_\parallel$ for pulsars with small predicted frequency
second derivatives. For pulsars with large predicted frequency second
derivatives, the dominant contributions to $\ddot{f}$ are thus
distinguishable and provide useful tools to probe the pulsar
parameters, in particular the radial velocity and the braking index. Below we discuss the pulsars in the first group of Fig.\,\ref{fig:f2terms} in an order of the potential magnitude of $\ddot{f}$. We compute the measurement uncertainty of $\ddot{f}$ by using Eqn.~\ref{eqn:sigf2} and assume no correlations in the timing noise.

Of the pulsars 
presented in Fig.\,\ref{fig:f2terms}, 
PSR\,J0437$-$4715 may
have the largest frequency second derivative. Due to its
significant proper motion, the value of $\ddot{f}_\parallel$ for
PSR\,J0437$-$4715 can easily exceed $10^{-29}$\,s$^{-3}$ when the
pulsar has a radial velocity $|\varv_\parallel|>12$\,km\,s$^{-1}$. We
note that for PSR\,J0437$-$4715 the local standard of rest velocity
already exceeds this limit ($\varv_\mathrm{LSR}=16.6$\,km\,s$^{-1}$,
see Fig.~\ref{fig:f2terms} and Table\,\ref{tab:ajf}). The
PSR\,J0437$-$4715 ephemeris by \citet{rhc16} achieves a timing
precision of $\sigma_\mathrm{rms}=0.3$\,$\upmu$s over a timespan of
$T=14.9$\,years. From the \citet{rhc16} ToAs, we determine a cadence
of $\Delta t=6.4$\,days, using 1.4\,GHz ToAs taken on different days.
With these timing ephemeris properties, one can achieve a measurement
accuracy of $\sigma(\ddot{f})=5.4\times10^{-30}$ s$^{-3}$ for the
frequency second derivative of PSR\,J0437$-$4715 (see
Eqn.~\ref{eqn:sigf2}). With this measurement uncertainty, a $5\sigma$ detection of $\ddot{f}$ would constrain the radial velocity of PSR\,J0437$-$4715 to $|\varv_\parallel|>33$\,km\,s$^{-1}$, as $\ddot{f}_\parallel$ dominates over the other contributions.

Though PSR\,J0437$-$4715 has a bright white dwarf binary companion,
the optical spectra of the white dwarf show no absorption lines due to
its low surface temperature \citep{dbv93,dkp12}. Hence, the radial
velocity of PSR\,J0437$-$4715 is unknown and pulsar timing may be the
only available method to measure it.

Radial velocity measurements from their binary companions are
available for the next two pulsars with possibly large $\ddot{f}$
values (Fig.\,\ref{fig:f2terms}). The first, PSR\,J1024$-$0719,
has a radial velocity of $\varv_\parallel=185\pm4$\,km\,s$^{-1}$
\citep{bjs16}, giving rise to $\ddot{f}_\parallel=3\times10^{-29}$
s$^{-3}$. As the PSR\,J1024$-$0719 binary orbit has an extremely long
period of $P_\mathrm{b}>200$\,yr \citep{bjs16,kkn16}, the orbit is not
modelled in the timing solution, giving rise to an observable
frequency second derivative of
$\ddot{f}=(-3.92\pm0.02)\times10^{-27}$\,s$^{-3}$ \citep{bjs16} due to
the gravitational effect of the remote binary companion. This is about
two orders-of-magnitude larger than the predicted
$\ddot{f}_\parallel$. Although the impact from the pulsar companion
pollutes the measurement of $\ddot{f}_\parallel$, the current
measurement uncertainty of $2\times10^{-29}$ s$^{-3}$ is comparable with 
the predicted value $3\times10^{-29}$ s$^{-3}$, therefore
$\ddot{f}_\parallel$ is important in the high precision timing of this
pulsar.

The second pulsar is PSR\,J1909$-$3744, for which the measured radial
velocity $\varv_\parallel=-73\pm30$\,km\,s$^{-1}$ (Table\,\ref{tab:psrv}) yields
a predicted value of $\ddot{f}_\parallel=(-8\pm3)\times10^{-30}$
s$^{-3}$. The current EPTA timing observations of this pulsar span 9.4 years
and have an excellent timing precision with rms residuals of 0.13\,$\upmu$s and a cadence of 8\,days \citep{dcl16}. Equation\,\ref{eqn:sigf2} thus predicts a measurement uncertainty for the frequency second derivative of $\sigma(\ddot{f})=2.5\times10^{-29}$\,s$^{-3}$. The timing baseline of PSR\,J1909$-$3744 was extended further in the recently released NANOGrav 11-year dataset by \citet{abb+18}. The longer timing baseline of 11.2 years, with a comparable rms timing residual of 0.15\,$\upmu$s, improves the measurement uncertainty to $\sigma(\ddot{f})=1.8\times10^{-29}$\,s$^{-3}$. Both data sets are not yet sensitive enough to measure $\ddot{f}_\parallel$. Given the current timing precision and cadence, a $5\sigma$ detection of $\ddot{f}_\parallel$ due to the $\varv_\parallel=-73$\,km\,s$^{-1}$ would require extending the timing baseline by another 11\,years, thereby decreasing the radial velocity uncertainty to 14\,km\,s$^{-1}$.

Two other pulsars may have $|\ddot{f}_\parallel|>10^{-29}$\,s$^{-3}$ if their radial velocities are larger than $|\varv_\parallel|>77$\,km\,s$^{-1}$ (for PSR\,J2124$-$3358) and 129\,km\,s$^{-1}$ (for PSR\,J1614$-$2230). The relatively poor timing precision of PSR\,J2124$-$3358 of $\sigma_{\rm rms}=3.2$\,$\upmu$s over a 9.4 year timing baseline \citep{dcl16} limits the probability of a significant detection of $\ddot{f}_\parallel$ and hence places poor constraints on its radial velocity. The situation is better for PSR\,J1614$-$2230, which has $\sigma_{\rm rms}=0.19$\,$\upmu$s over 5.3 years \citep{abb15}. A $5\sigma$ detection of $\ddot{f}_\parallel$ would require extending the timing baseline to 25\,years given the current timing precision and timing baseline, providing a measurement of radial velocities in excess of $|\varv_\parallel|>129$\,km\,s$^{-1}$.

The remaining four pulsars in the first group of Fig.~\ref{fig:f2terms} are PSRs J1721$-$2457, B1937+21, J1944+0907 and J2214+3000. As they have either a very large residual rms ($\sigma_{\rm rms}=11.7\ \upmu$s for PSR J1721$-$2457 and 34.5 $\upmu$s for PSR B1937+21\footnote{\cite{dcl16} obtained the residual rms of $\sigma_{\rm rms}=34.5\ \upmu$s for PSR B1937+21 by keeping the obvious structure in the residuals, which may be seen as the intrinsic feature of the pulsar. When the structure is mitigated, a smaller value of 
$\sigma_{\rm rms}=5.8\ \upmu$s \citep{rhc16} and even $1.5\ \upmu$s \citep{abb15} can be obtained.}, see \citealt{dcl16}) or a modest residual rms but with a short timing span ($\sigma_{\rm rms}=2.4\ \upmu$s and $T=5.8$ years for PSR J1944+0907 and 0.32 $\upmu$s and 2.1 years for PSR J2214+3000, see \citealt{abb15}), the current ability to measure the values of $\ddot{f}_\parallel$ in these pulsars is quite limited. 

However, as we can see from Fig.~\ref{fig:f2terms}, the relatively young pulsar PSR B1937+21 may have a significant and dominant intrinsic second derivative $\ddot{f}_0$. For a braking index of $n=3$, the value of $\ddot{f}_0$ reaches $8.8\times10^{-30}$\,s$^{-3}$. If the rms residuals can be reduced to $\sigma_{\rm rms}=0.3\,\upmu$s and the timing span can extend from the current $T=24$ to 35 years, then for a cadence of $\Delta t=10$ days, Eqn.~\ref{eqn:sigf2} gives us a measurement uncertainty of $\sigma(\ddot{f})=1.2\times10^{-30}$\,s$^{-3}$. It is therefore possible to measure $\ddot{f}_0$ when the braking index $|n|\ge2$ with a confidence level of 5$\sigma$ and a non-detection can constrain $|n|<2$. Although interpreting the strong value of $\ddot{f}_0$ of PSR B1937+21 may be problematic due to its significant and structural timing noise. \cite{scm13} attempted to explain the noise structure by assuming an asteroid belt around the pulsar. The gravitational potential from such a proposed asteroid belt may contribute to $\dot{f}$ and can not be modeled by our analyses, making the estimated $\dot{f}_0$ and consequently $\ddot{f}_0$ much larger than the real value. 

Given the observed radial velocities of the three remaining pulsars in Table~\ref{tab:psrv}, PSRs J1012+5307, J1738+0333 and J1903+0327 have small $\ddot{f}_\parallel$ of $(1.3\pm0.2)\times10^{-30}$, $(-1.3\pm0.5)\times10^{-31}$ and $(2.3\pm0.7)\times10^{-31}$\,s$^{-3}$, respectively. Since PSR J1012+5307 has been timed for $\sim$17 years with a precision of $\sigma_{\rm rms}=1.6\,\upmu$s, and PSR J1738+0333 has been observed for $\sim$7 years with $\sigma_{\rm rms}=3.0\,\upmu$s \citep{dcl16}, for a typical cadence of 10 days, the current data are not sensitive to the $\ddot{f}_\parallel$. Unless the timing precision is improved, extending the timing observations of these two pulsars for another 15 years will still not enable a measurement of the small $\ddot{f}_\parallel$. For PSR J1903+0327, since the value of $\ddot{f}_\parallel$ is not only small but also dominated by the contributions of $\ddot{f}_{\rm jerk}$ and $\ddot{f}_\perp$, the expectation of measuring $\ddot{f}$ due to its radial velocity is low.

\section{Conclusions}
In this paper we argue that the second derivative of the spin frequency, $\ddot{f}$, should be taken into account in the era of high-precision pulsar timing, as the timing baselines, precision and cadence have improved and will do so in the future. We derive the kinematic and dynamic contributions to the frequency second derivative $\ddot{f}$ in Eqn.~\ref{eqn:fullf2}. The $\ddot{f}_\parallel=3\varv_\parallel\mu^2f_0/c$ component can provide a novel approach to measure or constrain the radial velocity $\varv_\parallel$ of a pulsar, which is necessary to construct the three dimensional velocity and the trajectory of the pulsar in our Galaxy. The three dimensional velocity of pulsars will aid in understanding their formation and evolution. The \textsc{GALPY} package \citep{bov15} has been used to numerically compute the contributions to distinguish the impacts caused by the pulsar velocity from those by braking processes and the Galactic potential. Using the timing cadence, span and the rms residuals of ToAs, the measurement uncertainty $\sigma(\ddot{f})$ for the frequency second derivative has been derived in Eqn.~\ref{eqn:sigf2} and can be used to predict the detectability of $\ddot{f}$ by directly comparing the value of $\sigma(\ddot{f})$ and $|\ddot{f}|$. The analyses of $\sigma(\ddot{f})$ and components of $\ddot{f}$ in the pulsars monitored by EPTA, PPTA and NANOGrav have led to the conclusions below:

\begin{enumerate}

\item For some of the pulsars studied in this paper, the frequency second derivative induced by the radial velocity can have a large value ($|\ddot{f}_\parallel|\ge10^{-29}$\,s$^{-3}$) and may dominate the frequency second derivative if the radial velocity is moderately large ($|\varv_\parallel|\ge 50$\,km\,s$^{-1}$). If $\ddot{f}_\parallel$ dominates, then a measurement of $\ddot{f}$ can be used to detect the radial velocity, while a non-measurement can set an upper limit on the radial velocity.

\item Current long-term high precision timing datasets of some of the pulsars studied in this paper are on the brink of measuring the frequency second derivative induced by the radial velocity. For PSR J0437$-$4715, a $\ddot{f}$ would be measurable with a confidence level of 5$\sigma$ in the current data if the radial velocity $|\varv_\parallel|>33$\,km\,s$^{-1}$. If there is no detection by using the data from \cite{rhc16} then it may suggest that the radial velocity is less than 33\,km\,s$^{-1}$. For PSR J1024$-$0719, although the measured frequency second derivative is mainly caused by orbital motion, the measurement uncertainty of this value is already comparable with the predicted value of its $\ddot{f}_\parallel$. 

\item Continuing high precision timing of the PTA pulsars studied here would measure more $\ddot{f}_\parallel$ or impose stronger constraints on $\varv_\parallel$. For PSR J1909$-$3744, another sixteen-years timing at the current cadence and sensitivity would measure the $\ddot{f}_\parallel$ (5$\sigma$) and reduce the measurement uncertainty on $\varv_\parallel$ from 30 to 7\,km\,s$^{-1}$. 

\item The intrinsic frequency second derivative $\ddot{f}_0$ of PSR B1937+21 may be important. For a braking index of $n=3$, $\ddot{f}_0$ would be $8.8\times10^{-30}$\,s$^{-3}$ and dominate any observable $\ddot{f}$. If the rms residual of $\sigma_{\rm rms}=0.3\,\upmu$s and a timing span of $T=35$ years with a cadence of 10 days can be achieved, the $\ddot{f}_0$ induced by a braking index of $|n|\ge2$ can be measured (5$\sigma$), although the strong timing noise present in the timing data may hinder the measurement.

\end{enumerate}

\section*{Acknowledgements}
XJL acknowledges the support from the President's Doctoral Scholar Award from the University of Manchester. Pulsar research at Jodrell Bank Centre for Astrophysics is supported by a Consolidated Grant from the UK's Science and Technology Facilities Council. We thank A. Igoshev, M. Keith, J. McKee, B. Shaw and F. Verbunt for helpful discussions. This research made use of Astropy, a community-developed core Python package for Astronomy \citep{astropy2013}.

\appendix

\section{Derivatives of the unit position vector}
\label{apx:rdot}
The unit position vector of a pulsar at $\bmath{r}$ is $\hat{\bmath{r}}=\bmath{r}/r$, where $r=|\bmath{r}|$ is the distance to the pulsar. Considering the velocity $\bmath{\varv}$ and the acceleration $\bmath{a}$, the pulsar position is well approximated by $\bmath{r}=\bmath{r}_0+\bmath{\varv}t+\frac{1}{2}\bmath{a}t$, where $\bmath{r}_0$ is the position at the reference epoch and $t$ is a small temporal  lapse. Taylor-expanding $r$ up to $O(t^3)$, the unit vector becomes
\begin{equation}
\hat{\bmath{r}}=\frac{1}{r_0}\bigg[\bmath{r}_0+\bmath{\varv}_\perp t+\frac{1}{2r_0}\big(r_0\bmath{a}_\perp-2\varv_\parallel\bmath{\varv}_\perp-\varv_\perp^2\hat{\bmath{r}}_0\big)t^2\bigg]+O(t^3),
\end{equation}
where $\varv_\parallel=\bmath{\varv}\cdot\hat{\bmath{r}}_0$ and $\bmath{\varv}_\perp=\bmath{\varv}-\bmath{\varv}_\parallel$. At the reference epoch, differentiating $\hat{\bmath{r}}$ once gives $\dot{\hat{\bmath{r}}}=\bmath{\varv}_\perp/r_0$ and twice gives $\ddot{\hat{\bmath{r}}}=\big(r_0\bmath{a}_\perp-2\varv_\parallel\bmath{\varv}_\perp-\varv^2_\perp\hat{\bmath{r}}_0\big)/r_0^2$.

\section{Velocity transformation between LSRs}
\label{apx:LSRtrans}

The radial velocity of the LSR, $\varv_{\rm LSR}$, can be obtained in two steps. First, in the Galactic coordinate system, the relative velocity of the pulsar with respect to the Sun can be obtained by accounting for their peculiar velocities and the Galactic differential rotation (e.g. \citealt{vic17}). Here, we assume that the flat velocity curve of Galactic rotation applies for the Sun and the pulsars (e.g. \citealt{rmb14}). For the peculiar velocities, we use that from \cite{sbd2010} for the Sun and assume zero peculiar velocity for all the pulsars (i.e. pulsars rest in their LSRs). Next, using the velocity transformation between the equatorial and the Galactic coordinate system (e.g. \citealt{js87,bov11}) and the definition of Galactic coordinate system from \citet[section 1.5.3]{per97}, the pulsar $\varv_{\rm LSR}$ can be found and evaluated when given the distance and equatorial coordinate of the pulsar.

\section{Derivation of the measurement uncertainty}
\label{apx:sigmaf2}
The rotational parameters of a pulsar in a timing model, during which the best estimates are defined as
\begin{equation}
 \phi(t) = \phi(t_0)+f(t-t_0)+\frac{1}{2}\dot{f}(t-t_0)^2+\frac{1}{6}\ddot{f}(t-t_0)^3+\dots,
\end{equation}
where $\phi$ is the pulse phase and $t_0$ is an arbitrary reference epoch. To obtain a relation for the uncertainty on $\ddot{f}$, $\sigma(\ddot{f})$, we follow the general linear least squares derivation by \cite{ptv92}, where we write the timing model as a polynomial $\phi(x)=\sum_{k=1}^{4}\mathbfss{A}_k \mathbfss{X}_k(x)$ with $\mathbfss{A}=(\phi(t_0), f, \frac{1}{2}\dot{f}, \frac{1}{6}\ddot{f})$, $\mathbfss{X}=(1, x, x^2, x^3)$ and $x=t-t_0$. The inverse of the covariance matrix of the fitting parameters is given by 
\begin{equation}
\label{eqn:inverse}
\mathbfss{B}_{jk}=\sum_{i=1}^{N}\frac{\mathbfss{X}_j(x_i)\mathbfss{X}_k(x_i)}{\sigma^2_i}=\sum_{i=1}^N\frac{x_i^{j+k-2}}{\sigma_i^2},
\end{equation}
where $j, k$ ranges from 1 to 4, $N$ is the number of the data points and $\sigma_i$ is the measurement error of the $i$-th data point $x_i$. Therefore, the covariance between parameter $\mathbfss{A}_j$ and $\mathbfss{A}_k$ is ${\rm cov}(\mathbfss{A}_j, \mathbfss{A}_k)=(\mathbfss{B}^{-1})_{jk}$, and the variance of parameter $\mathbfss{A}_j$ is $\sigma^2(\mathbfss{A}_j)=(\mathbfss{B}^{-1})_{jj}$.

To connect the measurement uncertainty with the observational parameters, we assume that all measurements have identical uncertainty $\sigma$
and spread evenly over a timespan of $T$ at a cadence of $\Delta t$. Setting the reference epoch $t_0$ as the centre of the time span, the $i$-th data point is $x_i=i\Delta t$ with $i \in [-N/2, N/2]$, with $N=T/\Delta t$. Eqn.~\ref{eqn:inverse} then simplifies to:
\begin{equation}
\mathbfss{B}_{jk}=\frac{\Delta t^{j+k-2}}{\sigma^2}\sum_{i=-N/2}^{N/2}i^{j+k-2}=\frac{2\Delta t^{j+k-2}}{\sigma^2}\sum_{i=1}^{N/2}i^{j+k-2}.
\label{eqn:summation}
\end{equation}
Using Faulhaber's formula\footnote{e.g. see  Weisstein, Eric W. "Faulhaber's Formula." From MathWorld--A Wolfram Web Resource. http://mathworld.wolfram.com/FaulhabersFormula.html } to compute $\sum i^{j+k-2}$ in $\mathbfss{B}_{jk}$, inverting $\mathbfss{B}$ and considering $N\gg 1$, the covariance matrix becomes
\begin{equation}
{\rm cov}(\mathbfss{A}) = \begin{bmatrix}
  \frac{9\sigma^2\Delta t}{4T} & 0 & -\frac{15\sigma^2\Delta t}{T^3} & 0 \\
  0 & \frac{75\sigma^2\Delta t}{T^3} & 0 & -\frac{420\sigma^2\Delta t}{T^5} \\
  -\frac{15\sigma^2\Delta t}{T^3} & 0 & \frac{180\sigma^2\Delta t}{T^5} & 0\\
  0 & -\frac{420\sigma^2\Delta t}{T^5} & 0 & \frac{2800\sigma^2\Delta t}{T^7} \\
  \end{bmatrix}. 
\end{equation}
Hence, $\sigma^2(\mathbfss{A}_4)=2800\sigma^2\Delta t/T^7$, and with $\mathbfss{A}_4=\frac{1}{6}\ddot{f}$ and $\sigma=\sigma_{\rm rms}f$, we obtain $\sigma(\ddot{f})=6\sqrt{2800}f\sigma\Delta t^{1/2}T^{-7/2}$.

\bsp
\label{lastpage}
\bibliographystyle{mnras}

\begin{thebibliography}{}
\makeatletter
\relax
\def\mn@urlcharsother{\let\do\@makeother \do\$\do\&\do\#\do\^\do\_\do\%\do\~}
\def\mn@doi{\begingroup\mn@urlcharsother \@ifnextchar [ {\mn@doi@}
  {\mn@doi@[]}}
\def\mn@doi@[#1]#2{\def\@tempa{#1}\ifx\@tempa\@empty \href
  {http://dx.doi.org/#2} {doi:#2}\else \href {http://dx.doi.org/#2} {#1}\fi
  \endgroup}
\def\mn@eprint#1#2{\mn@eprint@#1:#2::\@nil}
\def\mn@eprint@arXiv#1{\href {http://arxiv.org/abs/#1} {{\tt arXiv:#1}}}
\def\mn@eprint@dblp#1{\href {http://dblp.uni-trier.de/rec/bibtex/#1.xml}
  {dblp:#1}}
\def\mn@eprint@#1:#2:#3:#4\@nil{\def\@tempa {#1}\def\@tempb {#2}\def\@tempc
  {#3}\ifx \@tempc \@empty \let \@tempc \@tempb \let \@tempb \@tempa \fi \ifx
  \@tempb \@empty \def\@tempb {arXiv}\fi \@ifundefined
  {mn@eprint@\@tempb}{\@tempb:\@tempc}{\expandafter \expandafter \csname
  mn@eprint@\@tempb\endcsname \expandafter{\@tempc}}}

\bibitem[\protect\citeauthoryear{{Antoniadis}}{{Antoniadis}}{2013}]{ant13}
{Antoniadis} J.~I.,  2013, PhD thesis, University of Bonn

\bibitem[\protect\citeauthoryear{{Antoniadis}, {van Kerkwijk}, {Koester},
  {Freire}, {Wex}, {Tauris}, {Kramer}  \& {Bassa}}{{Antoniadis}
  et~al.}{2012}]{avk+12}
{Antoniadis} J.,  {van Kerkwijk} M.~H.,  {Koester} D.,  {Freire} P.~C.~C.,
  {Wex} N.,  {Tauris} T.~M.,  {Kramer} M.,   {Bassa} C.~G.,  2012, \mn@doi
  [\mnras] {10.1111/j.1365-2966.2012.21124.x}, \href
  {http://adsabs.harvard.edu/abs/2012MNRAS.423.3316A} {423, 3316}

\bibitem[\protect\citeauthoryear{{Arzoumanian} et~al.,}{{Arzoumanian}
  et~al.}{2015}]{abb15}
{Arzoumanian} Z.,  et~al., 2015, \mn@doi [\apj] {10.1088/0004-637X/813/1/65},
  \href {http://adsabs.harvard.edu/abs/2015ApJ...813...65T} {813, 65}

\bibitem[\protect\citeauthoryear{{Arzoumanian} et~al.,}{{Arzoumanian}
  et~al.}{2018}]{abb+18}
{Arzoumanian} Z.,  et~al., 2018, preprint, \href
  {http://adsabs.harvard.edu/abs/2018arXiv180102617A} {} (\mn@eprint {arXiv}
  {1801.02617})

\bibitem[\protect\citeauthoryear{{Astropy Collaboration} et~al.,}{{Astropy
  Collaboration} et~al.}{2013}]{astropy2013}
{Astropy Collaboration} et~al., 2013, \mn@doi [\aap]
  {10.1051/0004-6361/201322068}, \href
  {http://adsabs.harvard.edu/abs/2013A%26A...558A..33A} {558, A33}

\bibitem[\protect\citeauthoryear{{Bassa} et~al.,}{{Bassa} et~al.}{2016}]{bjs16}
{Bassa} C.~G.,  et~al., 2016, \mn@doi [\mnras] {10.1093/mnras/stw1134}, \href
  {http://adsabs.harvard.edu/abs/2016MNRAS.460.2207B} {460, 2207}

\bibitem[\protect\citeauthoryear{{Beskin}}{{Beskin}}{1999}]{bes99}
{Beskin} V.~S.,  1999, \mn@doi [Physics Uspekhi]
  {10.1070/PU1999v042n11ABEH000665}, \href
  {http://adsabs.harvard.edu/abs/1999PhyU...42.1071B} {42, 1071}

\bibitem[\protect\citeauthoryear{{Blandford}, {Romani}  \&
  {Applegate}}{{Blandford} et~al.}{1987}]{bra87}
{Blandford} R.~D.,  {Romani} R.~W.,   {Applegate} J.~H.,  1987, \mn@doi
  [\mnras] {10.1093/mnras/225.1.51P}, \href
  {http://adsabs.harvard.edu/abs/1987MNRAS.225P..51B} {225, 51P}

\bibitem[\protect\citeauthoryear{{Bovy}}{{Bovy}}{2011}]{bov11}
{Bovy} J.,  2011, PhD thesis, New York University

\bibitem[\protect\citeauthoryear{{Bovy}}{{Bovy}}{2015}]{bov15}
{Bovy} J.,  2015, \mn@doi [\apjs] {10.1088/0067-0049/216/2/29}, \href
  {http://adsabs.harvard.edu/abs/2015ApJS..216...29B} {216, 29}

\bibitem[\protect\citeauthoryear{{Bovy}}{{Bovy}}{2017}]{bov17}
{Bovy} J.,  2017, \mn@doi [\mnras] {10.1093/mnras/stx1277}, \href
  {http://adsabs.harvard.edu/abs/2017MNRAS.470.1360B} {470, 1360}

\bibitem[\protect\citeauthoryear{{Callanan}, {Garnavich}  \&
  {Koester}}{{Callanan} et~al.}{1998}]{cgk98}
{Callanan} P.~J.,  {Garnavich} P.~M.,   {Koester} D.,  1998, \mn@doi [\mnras]
  {10.1046/j.1365-8711.1998.01634.x}, \href
  {http://adsabs.harvard.edu/abs/1998MNRAS.298..207C} {298, 207}

\bibitem[\protect\citeauthoryear{{Christian} \& {Loeb}}{{Christian} \&
  {Loeb}}{2015}]{cl15}
{Christian} P.,  {Loeb} A.,  2015, \mn@doi [\apj] {10.1088/0004-637X/798/2/78},
  \href {http://adsabs.harvard.edu/abs/2015ApJ...798...78C} {798, 78}

\bibitem[\protect\citeauthoryear{{Cordes} \& {Lazio}}{{Cordes} \&
  {Lazio}}{2002}]{cl01}
{Cordes} J.~M.,  {Lazio} T.~J.~W.,  2002, preprint, \href
  {http://adsabs.harvard.edu/abs/2002astro.ph..7156C} {} (\mn@eprint {arXiv}
  {0207156})

\bibitem[\protect\citeauthoryear{{Damour} \& {Taylor}}{{Damour} \&
  {Taylor}}{1991}]{dt91}
{Damour} T.,  {Taylor} J.~H.,  1991, \mn@doi [\apj] {10.1086/169585}, \href
  {http://adsabs.harvard.edu/abs/1991ApJ...366..501D} {366, 501}

\bibitem[\protect\citeauthoryear{{Damour} \& {Taylor}}{{Damour} \&
  {Taylor}}{1992}]{dt92}
{Damour} T.,  {Taylor} J.~H.,  1992, \mn@doi [\prd] {10.1103/PhysRevD.45.1840},
  \href {http://adsabs.harvard.edu/abs/1992PhRvD..45.1840D} {45, 1840}

\bibitem[\protect\citeauthoryear{{Danziger}, {Baade}  \& {della
  Valle}}{{Danziger} et~al.}{1993}]{dbv93}
{Danziger} I.~J.,  {Baade} D.,   {della Valle} M.,  1993, \aap, 276, 382

\bibitem[\protect\citeauthoryear{{Desvignes} et~al.,}{{Desvignes}
  et~al.}{2016}]{dcl16}
{Desvignes} G.,  et~al., 2016, \mn@doi [\mnras] {10.1093/mnras/stw483}, \href
  {http://adsabs.harvard.edu/abs/2016MNRAS.458.3341D} {458, 3341}

\bibitem[\protect\citeauthoryear{{Detweiler}}{{Detweiler}}{1979}]{det79}
{Detweiler} S.,  1979, \mn@doi [\apj] {10.1086/157593}, \href
  {http://adsabs.harvard.edu/abs/1979ApJ...234.1100D} {234, 1100}

\bibitem[\protect\citeauthoryear{{Durant}, {Kargaltsev}, {Pavlov}, {Kowalski},
  {Posselt}, {van Kerkwijk}  \& {Kaplan}}{{Durant} et~al.}{2012}]{dkp12}
{Durant} M.,  {Kargaltsev} O.,  {Pavlov} G.~G.,  {Kowalski} P.~M.,  {Posselt}
  B.,  {van Kerkwijk} M.~H.,   {Kaplan} D.~L.,  2012, \mn@doi [\apj]
  {10.1088/0004-637X/746/1/6}, \href
  {http://adsabs.harvard.edu/abs/2012ApJ...746....6D} {746, 6}

\bibitem[\protect\citeauthoryear{{Edwards}, {Hobbs}  \& {Manchester}}{{Edwards}
  et~al.}{2006}]{ehm06}
{Edwards} R.~T.,  {Hobbs} G.~B.,   {Manchester} R.~N.,  2006, \mn@doi [\mnras]
  {10.1111/j.1365-2966.2006.10870.x}, \href
  {http://adsabs.harvard.edu/abs/2006MNRAS.372.1549E} {372, 1549}

\bibitem[\protect\citeauthoryear{{Foster} \& {Backer}}{{Foster} \&
  {Backer}}{1990}]{fb90}
{Foster} R.~S.,  {Backer} D.~C.,  1990, \mn@doi [\apj] {10.1086/169195}, \href
  {http://adsabs.harvard.edu/abs/1990ApJ...361..300F} {361, 300}

\bibitem[\protect\citeauthoryear{{Freire} et~al.,}{{Freire}
  et~al.}{2011}]{fbw+11}
{Freire} P.~C.~C.,  et~al., 2011, \mn@doi [\mnras]
  {10.1111/j.1365-2966.2010.18109.x}, \href
  {http://adsabs.harvard.edu/abs/2011MNRAS.412.2763F} {412, 2763}

\bibitem[\protect\citeauthoryear{{Freire} et~al.,}{{Freire}
  et~al.}{2012}]{fwe12}
{Freire} P.~C.~C.,  et~al., 2012, \mn@doi [\mnras]
  {10.1111/j.1365-2966.2012.21253.x}, \href
  {http://adsabs.harvard.edu/abs/2012MNRAS.423.3328F} {423, 3328}

\bibitem[\protect\citeauthoryear{{Freire} et~al.,}{{Freire}
  et~al.}{2017}]{frk17}
{Freire} P.~C.~C.,  et~al., 2017, \mn@doi [\mnras] {10.1093/mnras/stx1533},
  \href {http://adsabs.harvard.edu/abs/2017MNRAS.471..857F} {471, 857}

\bibitem[\protect\citeauthoryear{{Hellings} \& {Downs}}{{Hellings} \&
  {Downs}}{1983}]{hd83}
{Hellings} R.~W.,  {Downs} G.~S.,  1983, \mn@doi [\apjl] {10.1086/183954},
  \href {http://adsabs.harvard.edu/abs/1983ApJ...265L..39H} {265, L39}

\bibitem[\protect\citeauthoryear{{Ho}}{{Ho}}{2015}]{ho15}
{Ho} W.~C.~G.,  2015, \mn@doi [\mnras] {10.1093/mnras/stv1339}, \href
  {http://adsabs.harvard.edu/abs/2015MNRAS.452..845H} {452, 845}

\bibitem[\protect\citeauthoryear{{Hobbs}, {Edwards}  \& {Manchester}}{{Hobbs}
  et~al.}{2006}]{hem06}
{Hobbs} G.~B.,  {Edwards} R.~T.,   {Manchester} R.~N.,  2006, \mn@doi [\mnras]
  {10.1111/j.1365-2966.2006.10302.x}, \href
  {http://adsabs.harvard.edu/abs/2006MNRAS.369..655H} {369, 655}

\bibitem[\protect\citeauthoryear{{Hogg}, {Blanton}, {Roweis}  \&
  {Johnston}}{{Hogg} et~al.}{2005}]{hbr+05}
{Hogg} D.~W.,  {Blanton} M.~R.,  {Roweis} S.~T.,   {Johnston} K.~V.,  2005,
  \mn@doi [\apj] {10.1086/431572}, \href
  {http://adsabs.harvard.edu/abs/2005ApJ...629..268H} {629, 268}

\bibitem[\protect\citeauthoryear{{Holmberg} \& {Flynn}}{{Holmberg} \&
  {Flynn}}{2004}]{hf04}
{Holmberg} J.,  {Flynn} C.,  2004, \mn@doi [\mnras]
  {10.1111/j.1365-2966.2004.07931.x}, \href
  {http://adsabs.harvard.edu/abs/2004MNRAS.352..440H} {352, 440}

\bibitem[\protect\citeauthoryear{{Jenet}, {Lommen}, {Larson}  \& {Wen}}{{Jenet}
  et~al.}{2004}]{jll04}
{Jenet} F.~A.,  {Lommen} A.,  {Larson} S.~L.,   {Wen} L.,  2004, \mn@doi [\apj]
  {10.1086/383020}, \href {http://adsabs.harvard.edu/abs/2004ApJ...606..799J}
  {606, 799}

\bibitem[\protect\citeauthoryear{{Johnson} \& {Soderblom}}{{Johnson} \&
  {Soderblom}}{1987}]{js87}
{Johnson} D.~R.~H.,  {Soderblom} D.~R.,  1987, \mn@doi [\aj] {10.1086/114370},
  \href {http://adsabs.harvard.edu/abs/1987AJ.....93..864J} {93, 864}

\bibitem[\protect\citeauthoryear{{Johnston} \& {Karastergiou}}{{Johnston} \&
  {Karastergiou}}{2017}]{jk17}
{Johnston} S.,  {Karastergiou} A.,  2017, \mn@doi [\mnras]
  {10.1093/mnras/stx377}, \href
  {http://adsabs.harvard.edu/abs/2017MNRAS.467.3493J} {467, 3493}

\bibitem[\protect\citeauthoryear{{Joshi} \& {Rasio}}{{Joshi} \&
  {Rasio}}{1997}]{jr97}
{Joshi} K.~J.,  {Rasio} F.~A.,  1997, \mn@doi [\apj] {10.1086/303916}, \href
  {http://adsabs.harvard.edu/abs/1997ApJ...479..948J} {479, 948}

\bibitem[\protect\citeauthoryear{{Kaplan} et~al.,}{{Kaplan}
  et~al.}{2016}]{kkn16}
{Kaplan} D.~L.,  et~al., 2016, \mn@doi [\apj] {10.3847/0004-637X/826/1/86},
  \href {http://adsabs.harvard.edu/abs/2016ApJ...826...86K} {826, 86}

\bibitem[\protect\citeauthoryear{{Keane}}{{Keane}}{2017}]{kea17}
{Keane} E.~F.,  2017, preprint, \href
  {http://adsabs.harvard.edu/abs/2017arXiv171101910K} {} (\mn@eprint {arXiv}
  {1711.01910})

\bibitem[\protect\citeauthoryear{{Khargharia}, {Stocke}, {Froning}, {Gopakumar}
   \& {Joshi}}{{Khargharia} et~al.}{2012}]{ksf+12}
{Khargharia} J.,  {Stocke} J.~T.,  {Froning} C.~S.,  {Gopakumar} A.,   {Joshi}
  B.~C.,  2012, \mn@doi [\apj] {10.1088/0004-637X/744/2/183}, \href
  {http://adsabs.harvard.edu/abs/2012ApJ...744..183K} {744, 183}

\bibitem[\protect\citeauthoryear{{Lazaridis} et~al.,}{{Lazaridis}
  et~al.}{2009}]{lwj09}
{Lazaridis} K.,  et~al., 2009, \mn@doi [\mnras]
  {10.1111/j.1365-2966.2009.15481.x}, \href
  {http://adsabs.harvard.edu/abs/2009MNRAS.400..805L} {400, 805}

\bibitem[\protect\citeauthoryear{{Lazio}}{{Lazio}}{2013}]{laz13}
{Lazio} T.~J.~W.,  2013, \mn@doi [Classical and Quantum Gravity]
  {10.1088/0264-9381/30/22/224011}, \href
  {http://adsabs.harvard.edu/abs/2013CQGra..30v4011L} {30, 224011}

\bibitem[\protect\citeauthoryear{{Lee}, {Wex}, {Kramer}, {Stappers}, {Bassa},
  {Janssen}, {Karuppusamy}  \& {Smits}}{{Lee} et~al.}{2011}]{lwk11}
{Lee} K.~J.,  {Wex} N.,  {Kramer} M.,  {Stappers} B.~W.,  {Bassa} C.~G.,
  {Janssen} G.~H.,  {Karuppusamy} R.,   {Smits} R.,  2011, \mn@doi [\mnras]
  {10.1111/j.1365-2966.2011.18622.x}, \href
  {http://adsabs.harvard.edu/abs/2011MNRAS.414.3251L} {414, 3251}

\bibitem[\protect\citeauthoryear{{Lorimer} \& {Kramer}}{{Lorimer} \&
  {Kramer}}{2004}]{lk04}
{Lorimer} D.~R.,  {Kramer} M.,  2004, {Handbook of Pulsar Astronomy}

\bibitem[\protect\citeauthoryear{{Lutz} \& {Kelker}}{{Lutz} \&
  {Kelker}}{1973}]{lk73}
{Lutz} T.~E.,  {Kelker} D.~H.,  1973, \mn@doi [\pasp] {10.1086/129506}, \href
  {http://adsabs.harvard.edu/abs/1973PASP...85..573L} {85, 573}

\bibitem[\protect\citeauthoryear{{Malov}}{{Malov}}{2017}]{mal17}
{Malov} I.,  2017, \mn@doi [\mnras] {10.1093/mnras/stx619}, \href
  {http://adsabs.harvard.edu/abs/2017MNRAS.468.2713M} {468, 2713}

\bibitem[\protect\citeauthoryear{{Matthews} et~al.,}{{Matthews}
  et~al.}{2016}]{mnf16}
{Matthews} A.~M.,  et~al., 2016, \mn@doi [\apj] {10.3847/0004-637X/818/1/92},
  \href {http://adsabs.harvard.edu/abs/2016ApJ...818...92M} {818, 92}

\bibitem[\protect\citeauthoryear{{Michel} \& {Li}}{{Michel} \&
  {Li}}{1999}]{ml99}
{Michel} F.~C.,  {Li} H.,  1999, \mn@doi [\physrep]
  {10.1016/S0370-1573(99)00002-2}, \href
  {http://adsabs.harvard.edu/abs/1999PhR...318..227M} {318, 227}

\bibitem[\protect\citeauthoryear{{Nan} et~al.,}{{Nan} et~al.}{2011}]{nlj11}
{Nan} R.,  et~al., 2011, \mn@doi [International Journal of Modern Physics D]
  {10.1142/S0218271811019335}, \href
  {http://adsabs.harvard.edu/abs/2011IJMPD..20..989N} {20, 989}

\bibitem[\protect\citeauthoryear{{Nice} \& {Taylor}}{{Nice} \&
  {Taylor}}{1995}]{nt95}
{Nice} D.~J.,  {Taylor} J.~H.,  1995, \mn@doi [\apj] {10.1086/175367}, \href
  {http://adsabs.harvard.edu/abs/1995ApJ...441..429N} {441, 429}

\bibitem[\protect\citeauthoryear{{Nice} et~al.,}{{Nice} et~al.}{2015}]{nds+15}
{Nice} D.,  et~al., 2015, {Tempo: Pulsar timing data analysis}, Astrophysics
  Source Code Library (\mn@eprint {ascl} {1509.002})

\bibitem[\protect\citeauthoryear{{Perryman}}{{Perryman}}{1997}]{per97}
{Perryman} M.~A.~C.,  1997, in {Bonnet} R.~M.,  et~al., eds,  ESA Special
  Publication Vol. 402, Hipparcos - Venice '97. pp~1--4

\bibitem[\protect\citeauthoryear{{Phinney}}{{Phinney}}{1992}]{phi92}
{Phinney} E.~S.,  1992, \mn@doi [Philosophical Transactions of the Royal
  Society of London Series A] {10.1098/rsta.1992.0084}, \href
  {http://adsabs.harvard.edu/abs/1992RSPTA.341...39P} {341, 39}

\bibitem[\protect\citeauthoryear{{Phinney}}{{Phinney}}{1993}]{phi93}
{Phinney} E.~S.,  1993, in {Djorgovski} S.~G.,  {Meylan} G.,  eds,
  Astronomical Society of the Pacific Conference Series Vol. 50, Structure and
  Dynamics of Globular Clusters. p.~141

\bibitem[\protect\citeauthoryear{{Prager}, {Ransom}, {Freire}, {Hessels},
  {Stairs}, {Arras}  \& {Cadelano}}{{Prager} et~al.}{2017}]{prf17}
{Prager} B.~J.,  {Ransom} S.~M.,  {Freire} P.~C.~C.,  {Hessels} J.~W.~T.,
  {Stairs} I.~H.,  {Arras} P.,   {Cadelano} M.,  2017, \mn@doi [\apj]
  {10.3847/1538-4357/aa7ed7}, \href
  {http://adsabs.harvard.edu/abs/2017ApJ...845..148P} {845, 148}

\bibitem[\protect\citeauthoryear{{Press}, {Teukolsky}, {Vetterling}  \&
  {Flannery}}{{Press} et~al.}{1992}]{ptv92}
{Press} W.~H.,  {Teukolsky} S.~A.,  {Vetterling} W.~T.,   {Flannery} B.~P.,
  1992, {Numerical recipes in FORTRAN. The art of scientific computing}

\bibitem[\protect\citeauthoryear{{Reardon} et~al.,}{{Reardon}
  et~al.}{2016}]{rhc16}
{Reardon} D.~J.,  et~al., 2016, \mn@doi [\mnras] {10.1093/mnras/stv2395}, \href
  {http://adsabs.harvard.edu/abs/2016MNRAS.455.1751R} {455, 1751}

\bibitem[\protect\citeauthoryear{{Reid} et~al.,}{{Reid} et~al.}{2014}]{rmb14}
{Reid} M.~J.,  et~al., 2014, \mn@doi [\apj] {10.1088/0004-637X/783/2/130},
  \href {http://adsabs.harvard.edu/abs/2014ApJ...783..130R} {783, 130}

\bibitem[\protect\citeauthoryear{{Sch{\"o}nrich}, {Binney}  \&
  {Dehnen}}{{Sch{\"o}nrich} et~al.}{2010}]{sbd2010}
{Sch{\"o}nrich} R.,  {Binney} J.,   {Dehnen} W.,  2010, \mn@doi [\mnras]
  {10.1111/j.1365-2966.2010.16253.x}, \href
  {http://adsabs.harvard.edu/abs/2010MNRAS.403.1829S} {403, 1829}

\bibitem[\protect\citeauthoryear{{Sesana}, {Vecchio}  \& {Colacino}}{{Sesana}
  et~al.}{2008}]{svc08}
{Sesana} A.,  {Vecchio} A.,   {Colacino} C.~N.,  2008, \mn@doi [\mnras]
  {10.1111/j.1365-2966.2008.13682.x}, \href
  {http://adsabs.harvard.edu/abs/2008MNRAS.390..192S} {390, 192}

\bibitem[\protect\citeauthoryear{{Shannon} et~al.,}{{Shannon}
  et~al.}{2013}]{scm13}
{Shannon} R.~M.,  et~al., 2013, \mn@doi [\apj] {10.1088/0004-637X/766/1/5},
  \href {http://adsabs.harvard.edu/abs/2013ApJ...766....5S} {766, 5}

\bibitem[\protect\citeauthoryear{{Shklovskii}}{{Shklovskii}}{1970}]{shk70}
{Shklovskii} I.~S.,  1970, \sovast, \href
  {http://adsabs.harvard.edu/abs/1970SvA....13..562S} {13, 562}

\bibitem[\protect\citeauthoryear{{Smits}, {Kramer}, {Stappers}, {Lorimer},
  {Cordes}  \& {Faulkner}}{{Smits} et~al.}{2009a}]{sks09}
{Smits} R.,  {Kramer} M.,  {Stappers} B.,  {Lorimer} D.~R.,  {Cordes} J.,
  {Faulkner} A.,  2009a, \mn@doi [\aap] {10.1051/0004-6361:200810383}, \href
  {http://adsabs.harvard.edu/abs/2009A%26A...493.1161S} {493, 1161}

\bibitem[\protect\citeauthoryear{{Smits}, {Lorimer}, {Kramer}, {Manchester},
  {Stappers}, {Jin}, {Nan}  \& {Li}}{{Smits} et~al.}{2009b}]{slk09}
{Smits} R.,  {Lorimer} D.~R.,  {Kramer} M.,  {Manchester} R.,  {Stappers} B.,
  {Jin} C.~J.,  {Nan} R.~D.,   {Li} D.,  2009b, \mn@doi [\aap]
  {10.1051/0004-6361/200911939}, \href
  {http://adsabs.harvard.edu/abs/2009A%26A...505..919S} {505, 919}

\bibitem[\protect\citeauthoryear{{Spitkovsky}}{{Spitkovsky}}{2004}]{spi04}
{Spitkovsky} A.,  2004, in {Camilo} F.,  {Gaensler} B.~M.,  eds,  IAU Symposium
  Vol. 218, Young Neutron Stars and Their Environments. p.~357 (\mn@eprint {}
  {astro-ph/0310731})

\bibitem[\protect\citeauthoryear{{Stairs}}{{Stairs}}{2003}]{sta03}
{Stairs} I.~H.,  2003, \mn@doi [Living Reviews in Relativity]
  {10.12942/lrr-2003-5}, \href
  {http://adsabs.harvard.edu/abs/2003LRR.....6....5S} {6, 5}

\bibitem[\protect\citeauthoryear{{Tauris} \& {Konar}}{{Tauris} \&
  {Konar}}{2001}]{tko01}
{Tauris} T.~M.,  {Konar} S.,  2001, \mn@doi [\aap]
  {10.1051/0004-6361:20010988}, \href
  {http://adsabs.harvard.edu/abs/2001A%26A...376..543T} {376, 543}

\bibitem[\protect\citeauthoryear{{Tong} \& {Kou}}{{Tong} \&
  {Kou}}{2017}]{tko07}
{Tong} H.,  {Kou} F.~F.,  2017, \mn@doi [\apj] {10.3847/1538-4357/aa60c6},
  \href {http://adsabs.harvard.edu/abs/2017ApJ...837..117T} {837, 117}

\bibitem[\protect\citeauthoryear{{Verbunt}, {Igoshev}  \& {Cator}}{{Verbunt}
  et~al.}{2017}]{vic17}
{Verbunt} F.,  {Igoshev} A.,   {Cator} E.,  2017, \mn@doi [\aap]
  {10.1051/0004-6361/201731518}, \href
  {http://adsabs.harvard.edu/abs/2017A%26A...608A..57V} {608, A57}

\bibitem[\protect\citeauthoryear{van Straten}{van Straten}{2003}]{van03a}
van Straten W.,  2003, PhD thesis, Swinburne University of Technology

\makeatother
\end{thebibliography}

\end{document}